\documentclass[12pt]{article}
\usepackage{bm,amsmath,epsf,epsfig,latexsym,amssymb,cite,lmodern}
\usepackage{graphics,float}
\usepackage[colorlinks]{hyperref}
\usepackage{ulem} 
\usepackage[utf8]{inputenc}

\newcommand{\nn}{\nonumber}

\newcommand{\lamc}{\Lambda_c(2595)^+}
\newcommand{\mcdd}{M_{CDD}}

\begin{document}

\thispagestyle{empty}

\title{Resonance on top of thresholds: the  $\Lambda_c(2595)^+$ as an extremely fine-tuned state}

\author{Zhi-Hui Guo$^{1,2,}$\footnote{Email address:
        \texttt{zhguo@mail.hebtu.edu.cn} },~
        J.~A.~Oller$^{3,}$\footnote{Email address:
        \texttt{oller@um.es} }\\[0.5em]
        {\it\small $^1$ Department of Physics, Hebei Normal University, Shijiazhuang 050024, People's Republic of China,}\\
        {\it\small $^2$ State Key Laboratory of Theoretical Physics, Institute of
Theoretical Physics, CAS,}\\
       {\it\small Beijing 100190, People's Republic of China,}\\
        {\it\small $^3$Departamento de F\'{\i}sica. Universidad de Murcia. E-30071 Murcia, Spain}}


\maketitle

\begin{abstract}
A dedicated study of the $\pi\Sigma_c$ scattering around its threshold is carried out in this work to probe the nature of  $\Lambda_c(2595)^+$.
We first demonstrate that the effective range expansion approach fails to work near the  $\Lambda_c(2595)^+$ pole position,
due to the presence of a nearby CDD pole around the $\pi\Sigma_c$ thresholds. We then develop a general framework to properly handle
the situation with a CDD pole accompanied by nearby thresholds, which is first elaborated for the single-channel case and then generalized to the coupled-channel study. 
The isospin-breaking effects of the three $\pi\Sigma_c$ channels with different thresholds are specially taken into account in our study. The finite-width effects from the $\Sigma_c$ baryons are considered and found to be relevant to give the  $\Lambda_c(2595)^+$ width fully compatible with its experimental value. 
Through the compositeness analysis, our robust conclusion is that the $\pi^0\Sigma_c^+$ component is  subdominant inside the  $\Lambda_c(2595)^+$.
\end{abstract}

\noindent{\bf PACS:}  14.20.Lq, 11.10.St, 03.65.Nk 
\\
\noindent{ {\bf Keywords:}  Charmed baryons. Meson-baryon scattering.   

%

\section{Introduction}
\label{sec:250715.1}

The recent discovery of a large number of new hadronic states, especially those with open or hidden heavy flavors, has triggered great interests both on the experimental and theoretical sides~\cite{150915.1.pdg}.
One of the common noticeable features of the newly observed hadrons is that many of them lie very close to the thresholds of two underlying intermediate states.
 It is then important to discriminate  that the observed peak structures from experimental analyses correspond to genuine resonances or threshold effects.
The resonance $\lamc$,  just lying on top of the threshold of $\pi\Sigma_c(2455)$ (simply denoted as $\pi\Sigma_c$ in the rest of this paper), is a typical kind of such states  and it is the focus of the current work.
However we should mention that our present formalism can be also straightforwardly generalized to other similar systems.

 The resonance $\lamc$ is a well established charm baryon, which was first observed by CLEO~\cite{Edwards:1994ar} and then confirmed later by other collaborations~\cite{Frabetti:1995sb,Albrecht:1997qa,Zhengjiu}.
The up-to-date experimental measurement is from the CDF collaboration~\cite{Aaltonen:2011sf}.
 Regarding its nature, the quark model can easily accommodate this state, by assigning symmetric orbital wave functions for the two constituent light quarks
 inside the $\lamc$~\cite{Copley:1979wj,Zhong:2007gp,Pirjol:1997nh,Tawfiq:1998nk,Zhu:2000py,Blechman:2003mq,Migura:2006ep}.
On the other hand,  a striking feature for the $\lamc$ is the 
 noticeable closeness of its mass to the $\pi\Sigma_c$ thresholds,  with quantum
numbers that are consistent with an $S$-wave $\pi\Sigma_c$ baryon resonance. This point
 has prompted several studies of this state by analyzing $\pi\Sigma_c$ scattering,
as well as including other relevant heavier intermediate states, such as $ND$, $ND^*$,
etc~\cite{Lutz:2003jw,Hofmann:2005sw,Geng:2014ina,Haidenbauer:2010ch,Ramos:2006vq,Ramos:2009vq,Nieves:2008dp,Nieves:2012hm,Nieves:2015jsa,Liang:2014kra}.
In these latter works, the unitarized chiral perturbation theory (UChPT) approach\footnote{Key references for this approach
are \cite{141215.1.kaiser.95,141015.1.jao.97,300915.2.ndorg,300915.3.plbkn,141215.2.pelaez.96,151215.3.oller.97,151215.4.oller.97}.} is extended from
 the strange baryons to the charm sector, mainly motivated by the similarity of the isoscalar $S$-wave $\Lambda(1405)$ and $\Lambda_c(2595)$
baryons~\cite{Hofmann:2005sw,Geng:2014ina,Haidenbauer:2010ch,Ramos:2006vq,Ramos:2009vq,Nieves:2008dp,Nieves:2012hm,Nieves:2015jsa,Liang:2014kra,Lutz:2003jw}.
The former state has a mass between the thresholds of $\pi\Sigma$ and $\bar{K}N$ and it also couples strongly to both channels~\cite{141215.1.kaiser.95,300915.3.plbkn,Guo:2012vv,141215.3.oset.98}.
The fact that the $\lamc$ lies  between the $\pi\Sigma_c$ and $DN$ thresholds
further supports the resemblance of the dynamics related to $\Lambda(1405)$ and $\Lambda_c(2595)$.
In Refs.~\cite{Lutz:2003jw,Geng:2014ina}, the $\lamc$ is generated by including the  channels $\pi\Sigma_c$ and $K\Xi_c$ (a much further away threshold)
and considered as a dynamically generated state by the nearby $\pi\Sigma_c$ channel.
Intermediate states including also the lightest charmed pseudoscalars are considered in Refs.~\cite{Hofmann:2005sw,Ramos:2006vq,Ramos:2009vq,Haidenbauer:2010ch}.
Based on the argument that the pseudoscalar charmed mesons $D$ and its vector companions $D^*$ should be treated equally according to the  heavy quark symmetry,
 channels including the latter ones are taken into account in Refs.~\cite{Nieves:2008dp,Nieves:2012hm,Nieves:2015jsa,Liang:2014kra}.
 These references stress the large couplings of the $ \lamc$ to both $DN$ and $D^*N$.

Nevertheless, we should mention that although the idea to generalize the UChPT study of $\Lambda(1405)$ to $\lamc$ is appealing, it is rather difficult for the previously mentioned UChPT studies
to precisely reproduce the mass and width of the
$\lamc$ simultaneously~\cite{Hofmann:2005sw,Geng:2014ina,Haidenbauer:2010ch,Ramos:2006vq,Ramos:2009vq,Nieves:2008dp,Nieves:2012hm,Nieves:2015jsa,Liang:2014kra,Lutz:2003jw}.
This may be due to the fact that $ND$ threshold is more than 200~MeV above the $\lamc$ mass, in contrast to the only 20~MeV or so difference between the $\bar{K}N$ threshold
 and the mass of $\Lambda(1405)$. Therefore in this work  we propose to make a delicate
 study for the $\lamc$ in the neighborhood energy region close to the $\pi\Sigma_c$ threshold, 
 for  $|\sqrt{s}-(m_{\pi}+m_{\Sigma_c})|\lesssim 5$~MeV, with $s$ the usual Mandelstam variable corresponding
to the center of mass energy squared of $\pi\Sigma_c$. Generally speaking, the effective range expansion (ERE) seems to offer an appropriate tool in this respect, due to the marked proximity of the $\lamc$
 and the $\pi\Sigma_c$ threshold. In fact, very recently the ERE has
been applied to investigate the near-threshold $S$-wave resonances in Refs.~\cite{150915.2.hyodo13,Long:2015pua}.
 As a matter of fact, the resonance $\Lambda_c(2595)$ was specially exemplified in these works as an ideal $S$-wave $\pi\Sigma_c$  baryon candidate.

 In the present study we include in the partial wave amplitudes 
the Castillejo-Dalitz-Dyson (CDD) poles~\cite{300915.6.cdd}, which effectively encode the extra degrees of freedom not directly
corresponding to the explicit $\pi\Sigma_c$ states, e.g. higher energy channels that are open much beyond  the  $\pi \Sigma_c$ threshold energy region, like  $ND, ND^*$ components, 
or compact quark-gluon states. Naively speaking the ERE should work well near the threshold region. But the convergence region can be severely restricted when a CDD pole lies close to
 threshold and this happens to be the situation for the $\lamc$, as shown in Sec.~\ref{sec.150915.1}. Therefore one should be cautious when applying the ERE up to the $\lamc$ pole
because the convergence of ERE becomes questionable, a point overlooked in Refs.~\cite{150915.2.hyodo13,Long:2015pua}.
We find a way out to this problem in the present work by explicitly keeping the CDD contributions around the threshold.   Furthermore, 
 because the mass of $\lamc$ lies just between the threshold of $\pi^0\Sigma_c^+$ and those of $\pi^+\Sigma_c^0$, $\pi^-\Sigma_c^{++}$, 
 a subtle issue that needs to be seriously taken into account is the isospin breaking effects associated with
the mass differences between the three $\pi\Sigma_c$ channels. This is one of the novelties of our present work, as isospin breaking effects are totally neglected in the previous  discussions on the  $\lamc$
~\cite{Hofmann:2005sw,Geng:2014ina,Haidenbauer:2010ch,Ramos:2006vq,Ramos:2009vq,Nieves:2008dp,Nieves:2012hm,Nieves:2015jsa,Liang:2014kra,Lutz:2003jw,150915.2.hyodo13,Long:2015pua}.

The paper is organized as follows. In Sec.~\ref{sec.150915.1}, we demonstrate in detail the problem of ERE in the $\lamc$ study.
Our single-channel solution to this problem is discussed in Sec.~\ref{sec.021015.1} and the coupled-channel formalism is developed in Sec.~\ref{sec.091015.1}. The finite-width effects from the $\Sigma_c$ baryons are considered in Sec.~\ref{sec.211015.1}. We then carry out the compositeness study of the $\lamc$ in Sec.~\ref{sec.151015.1} .
A short summary and our conclusions are given in Sec.~\ref{sec:sum}.

\section{General considerations for a resonance around thresholds and the ERE study of $\lamc$ }
\label{sec.150915.1}

The experimental values for the mass and width of the $\lamc$  are \cite{150915.1.pdg}
\begin{eqnarray}
\label{150915.1}
M_R=~2592.25\pm 0.28~{\rm MeV}~, \Gamma_R=~ 2.6\pm 0.6~ {\rm MeV}~.
\end{eqnarray}
These values should be compared with the threshold of  $\pi^0\Sigma_c^+$ (1), 2587.9~MeV, and the almost degenerate ones of
 $\pi^+\Sigma_c^0$ (2) and $\pi^-\Sigma_c^{++}$ (3), around 2593.5~MeV. Here we have indicated between parenthesis
 the numbering of every channel from lower to higher thresholds. As a result the $ \Lambda_c(2595)^+$ has the  appealing property of lying in between the thresholds of the lightest channel and those of the two heavier ones.
Notice that the differences between the $\lamc$ mass and the $\pi\Sigma_c$ thresholds are comparable to the  $\Sigma_c$ width,
which could lead to some non-negligible effects, as already noticed in Ref.~\cite{Blechman:2003mq}.
We shall also consider the small but finite widths from the $\Sigma_c$ states in this work.

Ref.~\cite{150915.2.hyodo13} proposes to study the $\Lambda_c(2595)^+$ through  $\pi\Sigma_c$ scattering in the isospin symmetric limit, so that common masses are used there
for the $ \Sigma_c$ and pions. Then an uncoupled $S$-wave ERE is employed as dynamical tool. However, one has to realize that there is an
ambiguity in taking definite values for the common isospin limit mass. The variation in the thresholds of the different channels is mainly due
 to the change from  the $\pi^+$ to the $\pi^0$ masses, since $M_{\pi^+}-M_{\pi^0}\simeq 4.5936(5)$~MeV while $M_{\Sigma_c^{0}}-M_{\Sigma_c^+}=0.9(4)$~MeV \cite{150915.1.pdg}.
When taking the $\pi^0\Sigma_c^+$ threshold as the isospin limit one, the $\lamc$ is around 4.4~MeV above it,
while when the $\pi^+\Sigma_c^0$ or $\pi^-\Sigma_c^{++}$ threshold is chosen instead, the $\lamc$ lies below it by around 1.2~MeV.
 As a result these two scenarios lead to dramatically different values for the ERE parameters (cf. Table~\ref{tab.220915.1}).
  This makes that an ERE in this case is extremely sensitive to the actual values taken for the masses  (a situation not realized indeed in Ref.~\cite{150915.2.hyodo13}).

In order to show this important point we make a necessary detour, since we propose to fix the
pole position of the $\Lambda_c(2595)^+$ to its actual physical value in Eq.~\eqref{150915.1},
independently of the values taken for the common isospin limit masses. 
 The fact that this procedure makes sense and is valuable for learning about the actual physical world is based on  the following two reasons.

Firstly, this proposal could be qualified as a gedankenexperiment in QCD because one has four active
free parameters in the energy range involved, namely, the common mass of $u$ and $d$ quarks,
the masses of the $s$ and $c$ quarks and the coupling constant (or $\Lambda_{QCD}$). 
 These free parameters would be fixed to the four conditions of choosing the common isospin masses of $\pi$ and $\Sigma_c$  together with the two extra constraints of reproducing the actual physical value of the pole position of the $\Lambda_c(2595)^+$ (that is, its mass and width).\footnote{Though the strange quark does not enter into our study directly, this does not necessarily mean that its effect is negligible. A typical example in this respect is the strange-quark contribution to the nucleon mass, which is found to range from -150 up to 0~MeV with large errors~\cite{Alarcon:2012nr}. Furthermore, in the present work we are talking about tiny changes in the thresholds of the states from changes in the masses of the constituent particles 
within the same isospin multiplets. Therefore it is plausible that in order to guarantee that these isospin limits are possible one should not exclude a priori the role of the strange quark.} This procedure would define a  possible QCD isospin limit hadronic world, that once it is properly characterized will give us valuable information on the actual physical situation.

Secondly, we can make use of the general principles for two-body scattering and show that
the parameters characterizing the interactions change little under variations in the chosen mass
within an isospin multiplet, while keeping the $\Lambda_c(2595)^+$ pole position fixed, as expected for isospin breaking corrections. Thus, by performing this gedankenexperiment we will
be able to determine rather approximately these parameters and calculate the actual partial-wave amplitudes, which in turn will provide us with extra relevant information, in particular,
regarding the $\Lambda_c(2595)^+$ resonance.

More specifically, due to the fact that the $\Lambda_c(2595)^+$ is almost on top of the thresholds of the $\pi\Sigma_c$ channels, with involved three-momenta much less than the pion mass,
we could take a pionless effective-field-theory point of view \cite{300915.1.pionless}, so that only local interactions enter in the dynamical picture of the process.
A scattering amplitude in this case has only right-hand cut or unitarity cut because, due to the contact nature of interactions,
 there are no crossed-channel cuts in the $\pi\Sigma_c$ scattering amplitude relevant for such low three momenta.
 The general expression for a partial wave when no crossed-channel cuts are present
 is given in Ref.~\cite{300915.2.ndorg}, making use of the N/D method \cite{051015.1.chew}.
Since the $\lamc$  lies so close to the $\pi \Sigma_c$ thresholds, it is very narrow and its quantum numbers correspond to a $\pi\Sigma_c$ S-wave resonance,
 to assume  only the  $\pi\Sigma_c$ elastic $S$-wave amplitude, $t(s)$, seems a safe assumption  (as also pointed out in  Ref.~\cite{150915.1.pdg}).
 This partial wave   can be expressed in such circumstances  as \cite{300915.2.ndorg,300915.3.plbkn}
(we consider here the single-channel case for simplicity and later we will generalize the discussion  to the coupled-channel case)
\begin{align}
\label{300915.1}
t(s)=\left[\sum_{i}\frac{\gamma_i^2}{s-M_{i,CDD}^2}+G(s)\right]^{-1}~.
\end{align}
Let us comment on the different elements appearing in the previous equation. Every term  $\gamma_i^2/(s-M_{i,CDD}^2)$  corresponds to the contribution of one CDD pole~\cite{300915.6.cdd}. In this way 
$t(s)$ is zero at $s=M_{i,CDD}^2$ since $1/t(s)$ has a pole at this point.
 As in the original paper \cite{300915.6.cdd} we concentrate here on
CDD poles lying on the real axis, so that both $M_{i,CDD}^2$ and $\gamma_i^2$ 
 are real parameters.
 The CDD poles are typically  associated with resonances and bound states because  $1/t(s)$
 for $s$ around $M_{i,CDD}^2$  crosses the pole associated with $\gamma_i^2/(s-M_{i,CDD}^2)$
 so that, if  the rest of contributions are smooth around $M_{i,CDD}^2$,  the real part of $1/t(s)$ would
 have also a zero not far from $M_{i,CDD}^2$. 
 For example, Dyson constructed  a model \cite{151215.5.dyson.56}
in which the relation between the CDD poles to bound states and resonances is explicitly exhibited.

The final term in Eq.~\eqref{300915.1} is the function $G(s)$ which
 is the scalar two-point loop function, or simply unitarity loop function, which comprises the unitarity cut, the only type of cut singularity in the present discussion.
This function can be expressed as \cite{300915.4.gfunc,300915.5.scalar}
\begin{align}
\label{300915.2}
G(s)=  \alpha(\mu^2)+
\frac{1}{(4\pi)^2}\left(  \log\frac{m_2^2}{\mu^2}-\varkappa_+\log\frac{\varkappa_+-1}{\varkappa_+}
-\varkappa_-\log\frac{\varkappa_--1}{\varkappa_-}
\right)~,
\end{align}
with
\begin{align}
\label{300915.3}
\varkappa_\pm=&\frac{s+m_1^2-m_2^2}{2s}\pm \frac{k}{\sqrt{s}}~,\nn\\
k=&\frac{\sqrt{(s-(m_1-m_2)^2)(s-(m_1+m_2)^2)}}{2\sqrt{s}}~,
\end{align}
being $k$ the modulus of the center-of-mass three-momentum  for a two-particle system with masses $m_1$ and $m_2$.
The constant $\alpha(\mu^2)$ in Eq.~\eqref{300915.2} is a subtraction constant,
with $\mu$ the renormalization scale. Notice that the combination of $\alpha(\mu^2)-\log\mu^2/(16\pi^2)$ is
independent of $\mu$. 

It can be easily verified from Eq.~\eqref{300915.1} that if the common isospin masses of the $\pi\Sigma_c$ are changed one can still keep the pole position
 of the $ \Lambda_c(2525)^+$ fixed at the physical value with little  changes of the parameters entering in Eq.~\eqref{300915.1}. We do this exercise  below in Sect.~\ref{sec.021015.1} by including one CDD pole.
 However, we point out that little change of the parameters in the partial-wave amplitude in Eq.~\eqref{300915.1} does not necessarily lead to similar results for the expansion parameters in the ERE approach.
In the following we explicitly show the extreme sensitivity of the ERE parameters, namely the scattering length and effective range,
to the actual values used for the $\pi\Sigma_c$ masses.

In this respect we  consider three possibilities by taking
 as common isospin masses the ones of every coupled channel separately, that is,
 $i)$  $M_{\Sigma_c}=M_{\Sigma_c^+}$, $M_\pi=M_{\pi^0}$,
$ii)$    $M_{\Sigma_c}=M_{\Sigma_c^0}$, $M_\pi=M_{\pi^+}$,
and $iii)$  $M_{\Sigma_c}=M_{\Sigma_c^{++}}$, $M_\pi=M_{\pi^-}$.
The choices $i)$ and $iii)$ represent the extreme cases of the  lowest and highest physical thresholds.
By now,  as in Ref.~\cite{150915.2.hyodo13}, we assume blindly that an ERE for $t(s)$  at the resonance pole position
is applicable, so that one can write
\begin{align}
\label{150915.2}
t(s)=& 8\pi\sqrt{s}\left(\frac{1}{a}+\frac{1}{2}r k^2-i k\right)^{-1}~,
\end{align}
where $a$ is the scattering length and $r$ is the effective range. These parameters are then fixed by imposing that Eq.~\eqref{150915.2} has a pole at $\sqrt{s}=M_R-i\Gamma_R/2$, cf. Eq.~\eqref{150915.1}.
 One then obtains
\begin{align}
\label{220915.2}
a=&\frac{2k_i}{|k_R|^2}~,\nn\\
r=&-\frac{1}{k_i}~,
\end{align}
 where  $k_r$ and $-k_i (k_i>0)$ are the real and imaginary parts of $k_R$, respectively, 
 the latter symbol being the value of $k$ at the pole position, $k_R=k_r-i k_i$ .  As a technical remark, we notice that the pole position 
 is located in the second Riemann sheet (RS), which implies that ${\rm Im}k<0$ (while in the physical
 or first Riemann sheet ${\rm Im}k>0$).

It is then clear from Eq.~\eqref{220915.2}  that there is a resonance only for $a>0$, $r<0$ and further it is required that
$a/2<-r$. For a more general discussion about the requirement on the $a$ and $r$ for the bound-, virtual- and resonance-state solutions, see Ref.~\cite{Ikeda:2011dx} for further details. The resulting numerical values for the different choices of common isospin masses are given in Table~\ref{tab.220915.1}, where the
 estimated uncertainties result from propagating the error bars in the $\Lambda_c(2595)^+$ pole position from Eq.~\eqref{150915.1}.
  Furthermore, since the pole position is very close  to threshold, for calculating these numbers we have used  nonrelativistic kinematics for $k$, 
\begin{align}
\label{220915.1}
k=&\sqrt{2\bar{\mu}(\sqrt{s}-M_{\Sigma_c}-M_\pi)}~,
\end{align}
with $\bar{\mu}=M_{\Sigma_c} M_\pi/(M_{\Sigma_c}+M_\pi)$  the reduced mass of the $\pi\Sigma_c$ system.

\begin{table}[ht]
\begin{center}
\begin{tabular}{l|ccccc}
\hline
Case & $M_{\Sigma_c}$ & $M_\pi$  & $M_{\Sigma_c}+M_{\pi}-M_R$~(MeV) & $a$~(fm) & $r$~(fm)  \\
\hline
$i)$  &  $M_{\Sigma_c^{+}}$ & $M_{\pi^0}$      & $-4.37$ & $1.66 \pm 0.38$  & $-40.1\pm 9.1$ \\
$ii)$   &  $M_{\Sigma_c^{0}}$ &  $M_{\pi^+}$   & $+1.06$  & $16.9 \pm 1.7$  &  $-10.4\pm 1.0$     \\
$iii)$   &  $M_{\Sigma_c^{++}}$ &  $M_{\pi^-}$ & $+1.30$  & $16.5\pm  1.4$  &  $-9.7\pm0.8$        \\
\hline
\end{tabular}
{\caption {\small  Single-channel analysis based on the ERE for $t(s)$, Eq.~\eqref{150915.2}.  The masses used for the $\Sigma_c$, $\pi$,
and the difference between the corresponding threshold and resonance mass are indicated in the 2nd, 3rd and
4th columns, in order. The values of the scattering length $a$ and effective range $r$
are given in the last two columns for each case, respectively. The error bars stem from the uncertainties in the mass and width of the $\lamc$ from Eq.~\eqref{150915.1}.  \label{tab.220915.1}}}
\end{center}
\end{table}

 It is noticeable the huge variations in the values of both $a$ and $r$, given in the 5th and 6th columns of Table~\ref{tab.220915.1}, respectively,  as the threshold  changes by just a few MeV. 
 Note that these variations arise from changes in the masses of the particles within the isospin multiplets. 
One can appreciate a change in $a$ by an order of magnitude and around a factor 4 for $r$ from cases $i)$ to $iii)$.
This is a clear indication of a fine tuning situation, and the results are extremely dependent on the exact values of the threshold. 
Let us stress that we are considering here only isospin breaking differences in the masses of the particles involved.
Another notorious fact from Table~\ref{150915.1} is the large magnitude of $r$, for case $i)$ it is actually huge, with values obtained that are
 much larger than $1/m_\pi$ or $1/\Lambda_{QCD}\simeq 1~{\rm fm}$.\footnote{Let us recall that $\Lambda_{QCD}\approx 190$~MeV with 4 quark flavors in the $\overline{MS}$ scheme \cite{150915.1.pdg}.}
 Indeed  pure potential scattering requires that the effective range should have a value around the range of interactions
\cite{021015.1.bethe,021015.2.preston}. From here it follows  the important conclusion that in order to account for the huge absolute values of $r$ one has to include another scale beyond the natural one for the range of the strong interactions.

There is indeed room in Eq.~\eqref{300915.1} to accomplish this by including a CDD pole near the threshold, so that the
new small energy scale would be the difference between the CDD mass and the threshold.
 The large magnitude of $r$ also indicates that the ERE could possibly have a very limited radius of convergence in $k$, much smaller than $m_\pi$.
This is actually related to the appearance of such a small new energy scale which is not required by branch points (due to the exchange of
particles) but due to the presence of a preexisting state that manifests in the need of including a CDD pole.
In this subtle situation it is then not obvious that one could apply the ERE to the $\Lambda_c(2595)^+$ because for the 4-MeV difference
between $M_R$ and the threshold in case $i)$ one would have $|k_R|\simeq 34$~MeV. Thus, one should consider the results in
Table~\ref{150915.1} as just indicative, and a more detailed discussion is necessary to
settle whether  the ERE could indeed be used at the $\Lambda_c(2595)^+$ pole position, as we do in the next section.

We point out that the changes in the values of the scattering lengths due to the different isospin masses
are much more dramatic than the large quark-mass dependence of the nucleon-nucleon scattering lengths, as calculated in
 Refs.~\cite{021015.3.qm.epelbaum,021015.4.qm.beane}, which is a paradigmatic example of fine-tuning problem. Here the situation, as remarked above, is much more impressive since we are just considering isospin breaking corrections in the masses while in the nucleon-nucleon case
 much larger changes in the pion masses  are involved, e.g. by approaching to the chiral limit.

\section{Single-channel CDD analysis}
\label{sec.021015.1}

Our main aim here is then to develop a picture based on Eq.~\eqref{300915.1} that explains the results in Table~\ref{tab.220915.1} and could be used even when ERE in Eq.~\eqref{150915.2} cannot be employed at the $\Lambda_c(2595)^+$ pole position.
  As we are just considering only one resonance it is then natural to include just one CDD pole. Then the $S$-wave $\pi \Sigma_c$ scattering  close to the threshold is given by
\begin{align}
\label{021015.1}
t(s)=\left[\frac{\gamma^2}{s-M_{CDD}^2}+G(s)\right]^{-1}~.
\end{align}
 We follow the convention of taking the renormalization scale $ \mu=M_{\pi^+}$ in the $G(s)$ function defined in Eq.~\eqref{300915.2}, since $M_\pi$ is an upper scale in  the low-energy effective field theory required for the scattering of $\pi\Sigma_c$ around the $\Lambda_c(2595)^+$ resonance. As a result we
 denote in the following the subtraction constant simply as $ \alpha$, without indicating its dependence on $\mu$.
 We now require that Eq.~\eqref{021015.1} reproduces for the cases $i)$ and $iii)$ the pole position of the $\Lambda_c(2495)^+$.
We do not explicitly consider case $ii)$ in what follows
  since its threshold is very close to that of $iii)$.
In this way we end with 4 equations, 2 for each case separately.

Instead of solving numerically these equations by using Eq.~\eqref{021015.1} we consider first
its nonrelativistic reduction  since it is simpler and allows us to derive algebraic expressions, which
are enlightening and  numerically accurate, given the proximity of the resonance to the threshold.
 In this limit Eq.~\eqref{021015.1} becomes simply
\begin{align}
\label{061015.2}
t(s)=&8\pi\sigma
\left[
\frac{\lambda^2}{\sqrt{s}-M_{CDD}}+\beta-ik
\right]^{-1}~,
\end{align}
where the new quantities $\lambda$ and $\beta$ are related to $\gamma$ and $ \alpha$, respectively, as
\begin{align}
\label{061015.3}
\lambda=&\gamma \sqrt{\frac{8\pi \sigma}{\sigma+M_{CDD}}}~,\\
\label{061015.3.b}
\beta=&8\pi\sigma \alpha +\frac{1}{\pi}\left( M_\pi\log\frac{M_\pi}{M_{\pi^+}}
+ M_{\Sigma_c}\log \frac{M_{\Sigma_c}}{M_{\pi^+}}\right)~,
\end{align}
with $\sigma=M_{\Sigma_c}+M_\pi$ and $k$ given by the Eq.~\eqref{220915.1}.
 Notice that the contribution from the first log on the right-hand side (rhs) of
the last equation  is negligibly small compared to the one from the second term.

Eq.~\eqref{061015.2} exhibits in a concise way how one can generate the results compiled
 in Table~\ref{tab.220915.1}, that are characterized by such large absolute values of $a$ and $r$,
as well as their dramatic variations with little changes in the thresholds.
 For that let us perform an expansion in  $k^2$ of  $k{\rm cot}\delta=8\pi\sigma/t(s)+i k$, with $ \delta$
the $S$-wave $\pi \Sigma_c$ phase shifts and  $t(s)$ given by nonrelativistic expression of Eq.~\eqref{061015.2}. We also use the nonrelativistic expression for $\sqrt{s}$,  which reads
\begin{align}
\label{061015.4}
\sqrt{s}=&\sigma+\frac{k^2}{2\bar{\mu}}~.
\end{align}
Then the following expansion results
\begin{align}
\label{061015.5}
k{\rm cot}\delta=&\frac{\lambda^2}{\sigma-M_{CDD}+k^2/2\bar{\mu}}+\beta\nn\\
=&\frac{\lambda^2}{\sigma-M_{CDD}}+\beta
-\frac{k^2 \lambda^2}{2\bar{\mu}(\sigma-M_{CDD})^2}
+\frac{\lambda^2}{\sigma-M_{CDD}}{\cal O}\left[ \left( \frac{k^2}{2\bar{\mu}(\sigma-M_{CDD})}\right)^n\right]~,
\end{align}
with $n>1$. From Eq.~\eqref{061015.5}, one can identify $1/a$ and $r$ as
\begin{align}
\label{061015.6}
a^{-1}=&\frac{\lambda^2}{\sigma-M_{CDD}}+\beta~,\nn\\
r=&-\frac{\lambda^2}{\bar{\mu}(\sigma-M_{CDD})^2}~.
\end{align}

It is then clear that in order to generate a large absolute value for $a$, one needs a strong cancellation between
the $\lambda^2/(\sigma-M_{CDD})$ and $\beta$. While to have a large magnitude for $r$, one would naturally expect $M_{CDD}\to \sigma$.
Equation \eqref{061015.6} also clearly shows why the ERE could fail to converge for values of $|k|^2\ll M^2_\pi$, since
it could perfectly be that $|\sigma-M_{CDD}|\ll M_\pi$ in the $\lamc$ case  because, as just discussed, we expect that
$M_{CDD}\approx \sigma$. As a result instead of applying the ERE in Eq.~\eqref{061015.5}, we consider directly Eq.~\eqref{061015.2}.

We now impose that $t(s)$ from Eq.~\eqref{061015.2}
has a pole at $s_R=(M_R-i\Gamma_R/2)^2$ in the second RS for masses corresponding
to channels $i)$ and $iii)$. In this RS Eq.~\eqref{061015.2} becomes  \cite{141015.1.jao.97}
\begin{align}
\label{261015.1}
t(s)^{II}=&8\pi\sigma
\left[
\frac{\lambda^2}{\sqrt{s}-M_{CDD}}+\beta+i k\right]^{-1}~,
\end{align}
 so that there is a change of sign in front of $k$ when comparing with Eq.~\eqref{061015.2}, where $k$ is calculated in the 1st RS with ${\rm Im}k>0$.
For a specific channel $j)$,  we denote by $\sigma_j$ the value of $\sigma$ and by $\rho_j$ the combination
\begin{align}
\label{071015.1}
\rho_j=\frac{1}{\pi}\left(M_{j,\pi}\log\frac{M_{j,\pi}}{M_{\pi^+}}+M_{j,\Sigma_c}\log\frac{M_{j,\Sigma_c}}{M_{\pi^+}}\right)~,
\end{align}
with $M_{j,\pi}$ and $M_{j,\Sigma_c}$ the corresponding masses of pion and $\Sigma_c$ in the channel $j)$.
Notice that $\rho_j$ is the last term on the rhs of Eq.~\eqref{061015.3.b}.
To require $t(s)^{II}$ in Eq.~\eqref{261015.1} to have a pole for both channels $i)$ and $iii)$ provides us four equations, which allow us
to determine $\gamma$, $\alpha$, $M_{CDD}^{i)}$ and $M_{CDD}^{iii)}$. Notice that we allow different CDD pole masses for every case since,
as it is clear from Eq.~\eqref{061015.6}, the final results are quite sensitive to $M_{CDD}$  as the threshold changes.
In contrast, one should expect smooth changes for the values of $\alpha$ and $\gamma$ and we take the same values for them in different channels. 
In addition we also distinguish between $k^{i)}_R$ and $k_R^{iii)}$ as $k$ depends on the threshold of the channel.

 These equations have two different solutions because they are quadratic in the CDD-pole masses.
In order to simplify the output for the solutions we take into account $M_{CDD}\approx \sigma$, so that
we take  $\lambda=\gamma\sqrt{4\pi}$ instead of  Eq.~\eqref{061015.3}. 
The algebraic solutions for $\alpha$, $\gamma^2$, $M_{CDD}^{i)}$ and $M_{CDD}^{iii)}$ read 
\begin{align}
\label{091015.1}
\alpha=&\frac{\Gamma_R\left(k_i^{iii)}-\rho_3\right)+2k_r^{iii)}\left(M_{CDD}^{iii)}-M_R\right)}{8\pi \sigma_3\Gamma_R}~,\nn\\
\gamma^2=&\frac{k_r^{iii)}\left[(M_{CDD}^{iii)}-M_R)^2+\Gamma_R^2/4\right]}{2\pi \Gamma_R}~,\nn \\
M_{CDD}^{i)}= &\frac{\chi_{11}\pm \sqrt{\chi_{12}}}{ k_r^{i)}  (k_r^{i)}\sigma_3^2-k_r^{iii)}\sigma_1^2)}~,
\nn\\
M_{CDD}^{iii)}= &\frac{\chi_{21} \pm \sqrt{\chi_{22}}}{ k_r^{iii)} (k_r^{i)}\sigma_3^2-k_r^{iii)}\sigma_1^2)}~,\nn\\
\chi_{11}=&M_R k_r^{i)}(k_r^{i)}\sigma_3^2-k_r^{iii)}\sigma_1^2)+ \Gamma_R k_r^{i)}\sigma_3(-k_i^{i)}\sigma_3+\rho_1\sigma_3+k_i^{iii)}\sigma_1
-\rho_3\sigma_1)/2~,\nn\\
\chi_{12}=&\frac{\Gamma_R^2}{4}k_r^{i)}k_r^{iii)}\sigma_1^2\left\{ 
\sigma_3^2\left[ k_r^{i)}(k_r^{i)}-k_r^{iii)})+(k_i^{i)}-\rho_1)^2\right]
-2\sigma_1\sigma_3(k_i^{iii)}-\rho_3)(k_i^{i)}-\rho_1)\right.\nn\\
+&\left.\sigma_1^2\left[k_r^{iii)}(k_r^{iii)}-k_r^{i)})+(k_i^{iii)}-\rho_3)^2\right]
\right\}~,\nn\\
 \chi_{21}=&M_R k_r^{iii)}(k_r^{i)}\sigma_3^2 - k_r^{iii)}\sigma_1^2)+ \Gamma_R k_r^{iii)} \sigma_1(-k_i^{i)}\sigma_3+\rho_1\sigma_3+k_i^{iii)}\sigma_1
-\rho_3\sigma_1)/2~,\nn\\
\chi_{22}=&\chi_{12}\frac{\sigma_3^2}{\sigma_1^2}~.
\end{align}

\begin{table}[ht]
\begin{center}
\begin{tabular}{lcccc}
\hline
                      & $\alpha$          & $\gamma$\,[MeV]         & $M^{i)}_{CDD}$\,[MeV] &  $M^{iii)}_{CDD}$\,[MeV] \\
\hline
1st solution       & $-0.03427(3)$ & $1.9(2)$              & $2592.3(2)$           & $2590.0(6)$                     \\
2nd solution         & $-0.03366(8)$ & $3.0(5)$             &  $2593.9(7)$        & $2596.2(7)$          \\
\hline
\end{tabular}
{\caption {\small From left to right, values for the parameters $\alpha$, $\gamma$, $M_{CDD}^{i)}$ and $M_{CDD}^{iii)}$
 after imposing that $t(s)$, Eq.~\eqref{021015.1}, 
has a pole at $s_R$   for channels $i)$ and $iii)$.  \label{tab.091015.1} } }
\end{center}
\end{table}

We point out that when deriving the expressions in Eq.~\eqref{091015.1}  a numerically small term proportional to $k^2$ from the expansion of $G(s)$ in
Eq.~\eqref{021015.1} is neglected. This is mainly done to derive the concise analytical results in Eq.~\eqref{091015.1}.
 Nevertheless, we mention that it is straightforward to keep the small $k^2$ term from the nonrelativistic expansion of $G(s)$. In this way
 one should add in the denominator of Eq.~\eqref{061015.2} the piece
\begin{equation}
\frac{k^2}{2\pi M_{j,\Sigma_c} M_{j,\pi}}
\bigg[ 2(M_{j,\Sigma_c}+M_{j,\pi}) - (M_{j,\Sigma_c}-M_{j,\pi})\log\frac{M_{j,\Sigma_c}}{M_{j,\pi}} \bigg]\,. 
\end{equation}

The values corresponding to Eq.~\eqref{091015.1} are almost the same as the exact solutions obtained by requiring that $t(s)$ given in Eq.~\eqref{021015.1} has a pole at $s_R$. 
This is because $k^2/2\bar{\mu}\ll \sqrt{s}$ and $M_{CDD}\approx \sigma$. 
We provide the exact solutions in Table~\ref{tab.091015.1}, where the two emerging solutions are distinguished. 
Notice that both  CDD pole masses
are very close to their respective thresholds, which  severely restricts the convergent radius of the ERE around $k=0$.
For channel $i)$ one has that $|k_R^{i)}|=34.2$~MeV, while the presence of the CDD pole implies that
the ERE does not converge for $|k|> 33.6$ and $39.3$~MeV for the first and second solutions, respectively.
The situation is similar for channel $iii)$ with $|k_R^{iii)}|=22.0$~MeV  but the ERE does not converge for
$|k|>30.6$ (1st solution) and $26.5$~MeV (2nd solution) due to  the proximity of $M_{CDD}^{iii)}$ to the threshold.
 Thus, one  can conclude from this analysis that
 the ERE is not an adequate tool to study the $\Lambda_c(2595)^+$ since
its convergence is  disrupted by the nearby CDD pole before reaching the pole position of the resonance.

\section{Coupled-channel CDD analysis I: stable asymptotic states}
\label{sec.091015.1}

We consider the generalization of Eq.~\eqref{021015.1} to the realistic $3\times 3$ coupled-channel scattering problem for the study of
the $\pi\Sigma_c$-threshold energy region  where the $\Lambda_c(2595)^+$ sits.  We first discuss the results obtained
with stable asymptotic states, that is, taking zero widths for the $\Sigma_c$ baryons
and afterwards we estimate in Sec.~\ref{sec.211015.1} the effects of including  finite $\Sigma_c$ widths. Our preferred outcomes
correspond to the 1st solution in Table~\ref{tab.091015.1} because,  as shown below,  
in the $3\times 3$ coupled-channel case they  give rise to a resonance signal in accordance with the mass and width of the
 $\Lambda_c^+(2595)$.

\subsection{Scattering equation}
\label{sec.201015.1}

 To end with the adequate coupled-channel equation let us rewrite Eq.~\eqref{021015.1} as an algebraic Bethe-Salpeter equation \cite{141015.1.jao.97}
\begin{align}
\label{141015.1}
t(s)=w(s)-w(s) G(s) t(s)~,
\end{align}
with $w(s)$ the inverse of the CDD pole contribution, namely,
\begin{align}
\label{141015.2}
w(s)=&\frac{s-M_{CDD}^2}{\gamma^2}~.
\end{align}
E.g. the isoscalar and scalar $\pi\pi$ partial wave amplitude at
leading order in Chiral Perturbation Theory has precisely this form \cite{300915.2.ndorg}.

 Now, let us discuss the generalization of Eqs.~\eqref{141015.1} and \eqref{141015.2}  to the coupled-channel case under consideration.
Since the $\Lambda_c^+(2595)$ is an isoscalar resonance \cite{150915.1.pdg}
there is an extra factor 1/3 multiplying $w(s)$ in Eq.~\eqref{141015.2} for each transition matrix element (as it is also clear from the Wigner-Eckart theorem).
We denote  this contribution as
\begin{align}
\label{141015.3}
{\cal K}_{ij}(s)=\frac{1}{3}w(s)=\frac{s-M_{CDD}^2}{3\gamma^2}~,
\end{align}
with $i,j,=1,~2,~3.$ Here we are taking common values for the CDD pole and $\gamma$ in all three channels, so that isospin symmetry is preserved for these
matrix elements.  The main isospin breaking corrections between the different coupled channels  are expected to arise from the fact of using specific scalar loop functions
$G_i(s)$,  $i=1,2,3$, Eq.~\eqref{300915.2}, for every channel 
   due to the associated branch point singularity at each nearby threshold.
    One could argue about different CDD pole masses for different channels but then the
expression for the matrix  elements ${\cal K}_{ij}(s)$ would become ambiguous, and moreover our results are phenomenologically suited, as we discuss below.
In addition,  the changes in the single-channel case  of $M_{CDD}$ of $\sim 2\pm 1$~MeV for channels $i)$ and $iii)$ in both the 1st and 2nd solutions,
see Table~\ref{tab.091015.1}, seem to indicate  that isospin breaking effects in $M_{CDD}$
are expected to be  similar to the small mass difference within the $\Sigma_c$ multiplet, rather than to the much larger
differences in $M_\pi$ (which are the main sources for the variation in the $\pi\Sigma_c$ thresholds, as
pointed out above).

We can then generalize the single-channel formalism in Eq.~\eqref{141015.1} to the coupled-channel case by prompting $t(s)$ as the matrix,
\begin{align}
\label{141015.4}
&t(s)={\cal K}(s)-{\cal K}(s)G(s) t(s)~,\nn\\
&\left[I+{\cal K}(s)G(s)\right] t(s) ={\cal K}(s)~,
\end{align}
where we denote by $G(s)$, ${\cal K}(s)$ and $t(s)$ the $3\times 3$ matrices with matrix elements $G_i(s)$, ${\cal K}_{ij}(s)$ and $t_{ij}(s)$, in order.
Note that the matrix $G(s)$ is diagonal. The  solution of Eq.~\eqref{141015.4}  can be recast in the following form
\begin{align}
\label{141015.5}
t(s)=\left[I+{\cal K}(s) G(s)\right]^{-1}{\cal K}(s)~.
\end{align}
Indeed, all the matrix elements $t_{ij}(s)$ resulting from the previous equation are the same, due to the
form of  ${\cal K}(s)$ given in Eq.~\eqref{141015.3}, and the explicit expressions for $t_{ij}(s)$  read
\begin{align}
\label{201015.1}
t_{ij}(s)=&
\left[\frac{3 \gamma^2}{s-M_{CDD}^2}+ G_1(s)+G_2(s)+G_3(s) \right]^{-1}~.
\end{align}
In terms of Eq.~\eqref{141015.3}, it is not difficult to understand Eq.~\eqref{201015.1} since the interaction between
all the three channels is driven by the same function ${\cal K}_{ij}(s)$.  We now have analogous expressions to  Eq.~\eqref{021015.1}
for $t_{ij}(s)$ but with a sum over the three possible intermediate states.

 Differences between Eqs.~\eqref{021015.1} and ~\eqref{201015.1}  arise because in the later  equation 
we employ the physical masses of the three $\pi\Sigma_c$ channels in the different $G_i(s)$ and in addition $M_{CDD}$ in the coupled-channel case does not
correspond a priori to any of the single-channel determinations in Table~\ref{tab.091015.1}.
While for the other parameters,  $\alpha$ and $\gamma$,  we take their values from the single-channel analysis of Sec.~\ref{sec.021015.1},
 whose  explicit numbers are also provided in Table~\ref{tab.091015.2} for later convenience. 
   Then, at this stage our only free parameter is $M_{CDD}$.

\begin{table}
\begin{center}
\begin{tabular}{lccccccc}
\hline
                  & $\alpha$ & $\gamma$\,[MeV]         & $M_{CDD}$\,[MeV] & $M_{CDD}$\,[MeV] &   $M_R$\,[MeV] & $\Gamma_R$\,[MeV]  &   $M_{CDD}$\,[MeV] \\
 RS                 &                &                                     & $(1,1,1)$              & $(1,0,0)$              &   $(1,0,0)$    & $(1,0,0)$                                 &         $(1,0,0)$          \\
\hline
1st sol.          & $-0.03427(3)$ & $1.9(2)$              & $2590.9(7)$  &   $2594.7(6)$ & $2592.3(3)$ & $2.3(3)$   & $2594.2(5)$ \\
2nd sol.           & $-0.03366(8)$ & $3.0(5)$             & $2595.0(2)$  &  $2594.3(5)$ & $2592.6(4)$ & $0.9(1)$   & $2594.4(5)$ \\
\hline
\end{tabular}
{\caption {\small (From left to right.)  The first two parameters ($\alpha$ and $\gamma$)
 were determined by the single-channel analysis in Sec.~\ref{sec.021015.1}. The corresponding RS where the pole lies is indicated in the second row.
  The fourth and fifth  columns give the CDD pole mass for the $3\times 3$ coupled-channel analysis with asymptotic stable states, while the last
three columns  refer to the coupled-channel analysis
including finite widths for $ \Sigma_c$. The resonance mass ($M_R$) and half width ($\Gamma_R/2$) refer to the real and imaginary parts of the pole position in the complex energy plane respectively, and the CDD-pole mass is reported in the last column for the finite-width case. 
\label{tab.091015.2}}}
\end{center}
\end{table}

\subsection{Application to the case of stable asymptotic states}
\label{sec.201015.2}

We first consider the limit of zero width for the $\Sigma_c$ baryons which have indeed small widths.
  The updated PDG \cite{150915.1.pdg} gives the averages $\Gamma_{\Sigma_c^{++}}=1.89^{+0.09}_{-0.18}$~MeV,
 $\Gamma_{\Sigma_c^0}=1.83^{+0.11}_{-0.19}$~MeV  while
 for the $\Sigma_c^+$  only an  upper bound is provided, $\Gamma_{\Sigma_c^+}<4.6$~MeV.

The value of $M_{CDD}$ can be fixed by requiring that  $t_{ij}(s)$, Eq.~\eqref{201015.1}, has a pole at $s_R$. However,
for the coupled-channel case one has to specify the nonphysical RS in which this pole lies. For the present problem, with channels 2 ($\pi^+\Sigma_c^0$) and 3 ($\pi^-\Sigma_c^{++}$) almost degenerate, we have two main RS's that connect continuously with the physical one. We denote these RS's by (1,0,0) and (1,1,1)
or by 2nd and 3rd RS's, respectively. The former connects continuously with the physical RS in the energy region between the thresholds of channels 1 and 2,
that is, for $M_{\pi^0}+M_{\Sigma_c^{+}}<\sqrt{s}<M_{\pi^+}+M_{\Sigma_c^0}$, while the latter does it above the threshold for channel 3,  namely,
$\sqrt{s}>M_{\pi^-}+M_{\Sigma_c^{++}}$.

To perform the analytical continuation to the nonphysical RS's we employ the procedure of Ref.~\cite{141015.1.jao.97} so that the unitarity loop function
$G_j(s)$ in its associated nonphysical RS,\footnote{For $n$-coupled channels there are $2^n$ RS's,  with two sheets associated to each channel.}
$G_j^{II}(s)$, is given by
\begin{align}
\label{201015.2}
G_j^{II}(s)=&G_j(s)+ i\frac{k_j(s)}{4\pi\sqrt{s}}~,
\end{align}
where $k_j(s)$ is the three-momentum for channel $j$ and it is required that ${\rm Im}k_j(s)>0$.
In this way, the $t_{ij}(s)$ in the (1,0,0) RS  reads
\begin{align}
\label{201015.3}
t^{II}_{ij}(s)=&
\left[\frac{3 \gamma^2}{s-M_{CDD}^2}+ G_1^{II}(s)+G_2(s)+G_3(s) \right]^{-1}~,
\end{align}
and for the (1,1,1) RS it becomes
\begin{align}
\label{201015.4}
t^{III}_{ij}(s)=&
\left[\frac{3 \gamma^2}{s-M_{CDD}^2}+ G_1^{II}(s)+G_2^{II}(s)+G_3^{II}(s) \right]^{-1}~.
\end{align}

\begin{figure}[H]
\begin{center}
\begin{tabular}{ccc}
  \includegraphics[width=.3\textwidth]{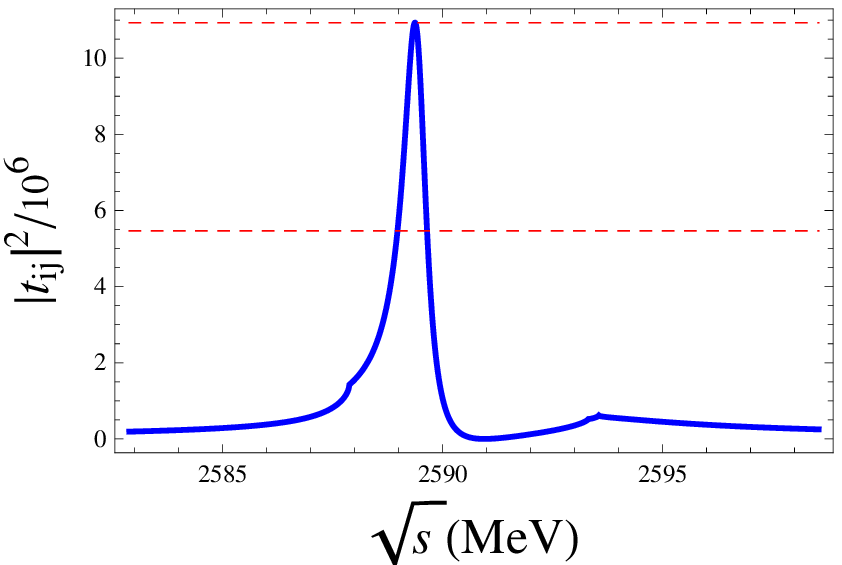} &
  \includegraphics[width=.3\textwidth]{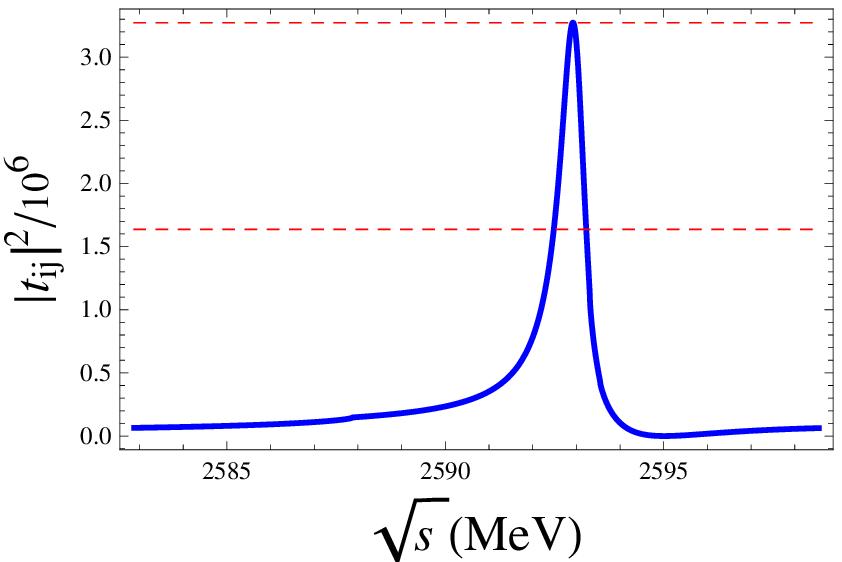}&
  \includegraphics[width=.3\textwidth]{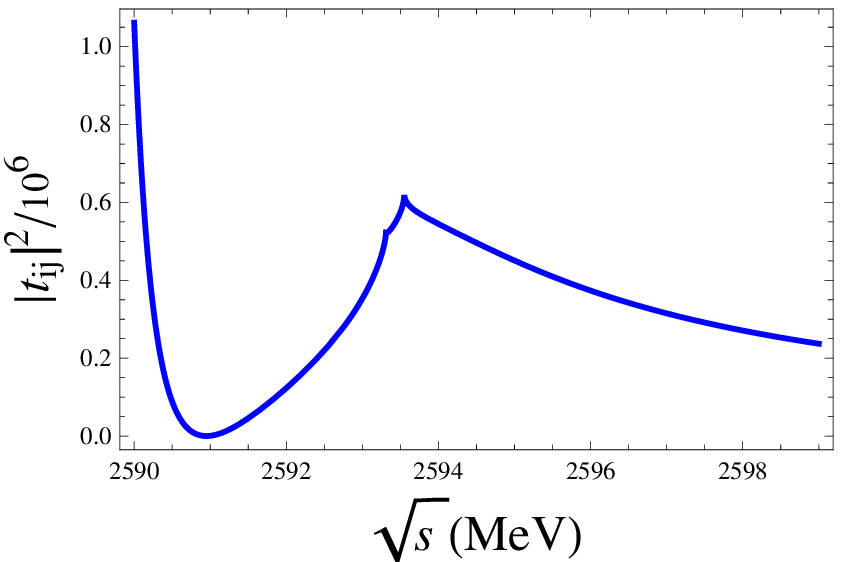}
\end{tabular}
\caption{{\protect{\small The partial wave $|t_{ij}(s)|^2$ is plotted along the physical axis in all the panels (ordered from left to right) in the case of asymptotic stable states.
 In the first and second panels we consider the physical axis within an interval of $\pm 5$~MeV above and below the highest and lightest thresholds  for
the 1st and 2nd solutions, respectively.  We also indicate by horizontal lines the maximum value at the peak and its half value. The most right panel corresponds to 
the 1st solution and the region around  the two heavier thresholds is highlighted.
\label{fig.211015.1}}}}
\end{center}
\end{figure}

It is possible to adjust $M_{CDD}$ so as to reproduce the physical parameters attached to the pole of the $\Lambda_c^+(2595)$
in the RS (1,1,1) for both the 1st and 2nd solutions at the one-sigma level within the experimental uncertainties.
  Notice that this condition implies two equations which are not trivial to be fulfilled
 since we have only one free parameter at our disposal.
 We then obtain  the values for $M_{CDD}$ given in the third column of Table~\ref{tab.091015.2} for the 1st and 2nd solutions.
 In particular the central values given in Table~\ref{tab.091015.2}
imply a pole in the (1,1,1) RS located at  $2592.1-i\,1.8$~MeV for the 1st solution and at $2592.3-i 1.6~$MeV for the 2nd one.

 However, notice that the (1,1,1) RS only connects continuously with the physical RS for $\sqrt{s}>M_{\pi^+}+M_{\Sigma_c^{++}}=2593.5$~MeV,
while the poles are below this threshold. It happens indeed that  along the physical energy axis $|t_{ij}(s)|^2$ does have a resonance behavior but it does not correspond to the experimentally determined
parameters of the $\Lambda_c^+(2595)$, Eq.~\eqref{150915.1}. To show this we plot $|t_{ij}(s)|^2$ along the physical axis  in  Fig.~\ref{fig.211015.1}, within an interval in $\sqrt{s}$ from $M_{\pi^0}+M_{\Sigma_c^{+}}-5$~MeV up to $M_{\pi^-}+M_{\Sigma_c^{++}}+5$~MeV.
  For the 2nd solution (2nd panel from left to right) there is a clear resonance structure just below the two heavier thresholds at around 2593~MeV, close to the mass of the $\Lambda_c^+(2595)$, but its width, of only 0.6~MeV, is much smaller than that of the $\Lambda_c^+(2595)$.
For the 1st solution (1st panel) the situation is similar, although the peak lies at lower energies, at around 2589.5~MeV. In the most right panel 
 we show closely the region around the $\pi^+\Sigma_c^0$ and $\pi^-\Sigma_c^{++}$ thresholds for the 1st solution and one cannot appreciate any
resonance behavior but just a cusp effect due to the opening of the thresholds. Note also the presence of zeroes in $|t_{ij}(s)|^2$ at $\sqrt{s}=M_{CDD}$ for
every solution, as it corresponds to the CDD pole.

We point out that the resonance structures along the real axis shown in Fig.~\ref{fig.211015.1} correspond to poles of $t_{ij}(s)$ in the (1,0,0) RS,
instead of those in the (1,1,1) RS.
The poles in the (1,0,0) RS are found to be located at  $2589.5-i\,0.3$~MeV  (1st sol.) and $2593.0-i\,0.4$~MeV (2nd sol.), which are consistent with peaks shown in the first two 
panels in Fig.~\ref{fig.211015.1}. The RS (1,0,0) is the one that connects continuously with the physical axis between the thresholds of $\pi^0\Sigma_c^+$ and $\pi^+\Sigma_c^0$,
along which the resonance signal occurs.  Interestingly, the two poles at the (1,0,0) and (1,1,1) RS's are connected by changing continuously between these RS's.  This can be explicitly verified by introducing a continuous parameter $\nu\in [0,1] $,
 such that the  functions $G_j(s)^{II}$ in  Eq.~\eqref{201015.2} for channels 2 and 3 are replaced by
\begin{eqnarray} 
\label{211015.2}
G_j^{II}(s) &\to &G_j(s)+ i\nu \frac{k_j(s)}{4\pi\sqrt{s}}~,~j=2,\, 3.
\end{eqnarray}
In this way $\nu=0$ corresponds to the RS (1,0,0) and $\nu=1$ to RS (1,1,1). Then, one can observe how one pole evolves into the other. 

\begin{figure}[H]
\begin{center}
\begin{tabular}{cc}
  \includegraphics[width=.45\textwidth]{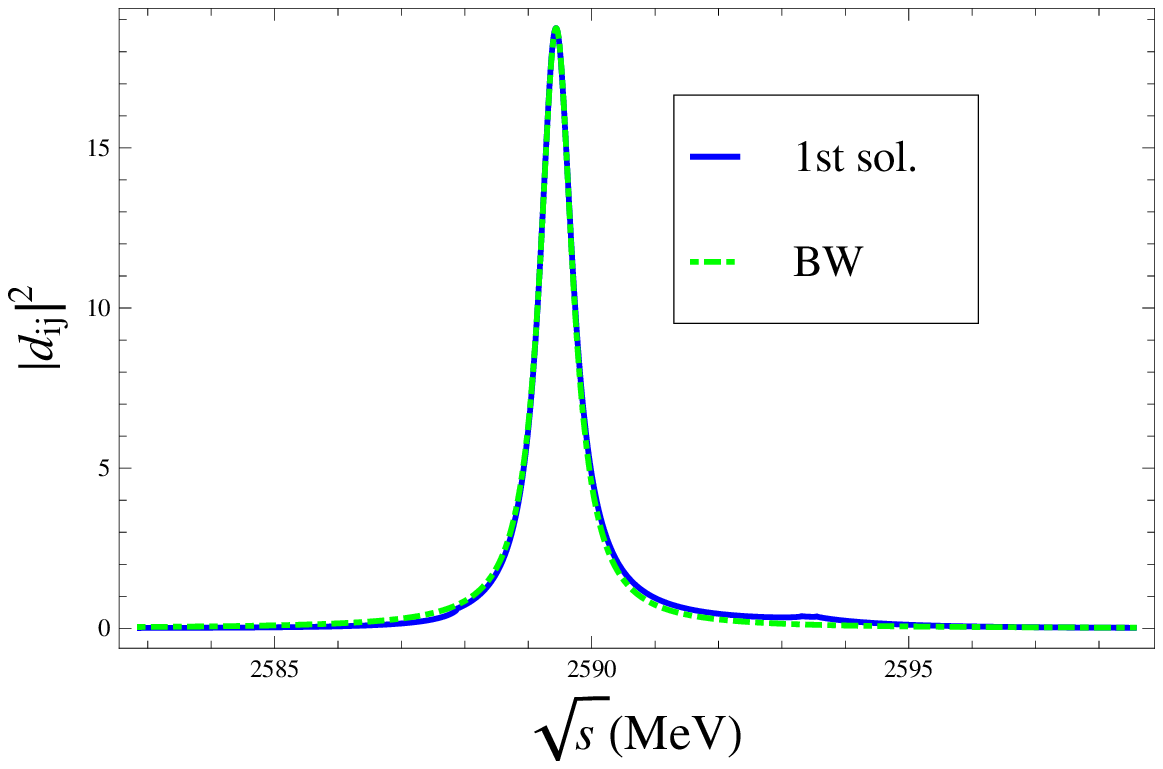}
 & \includegraphics[width=.45\textwidth]{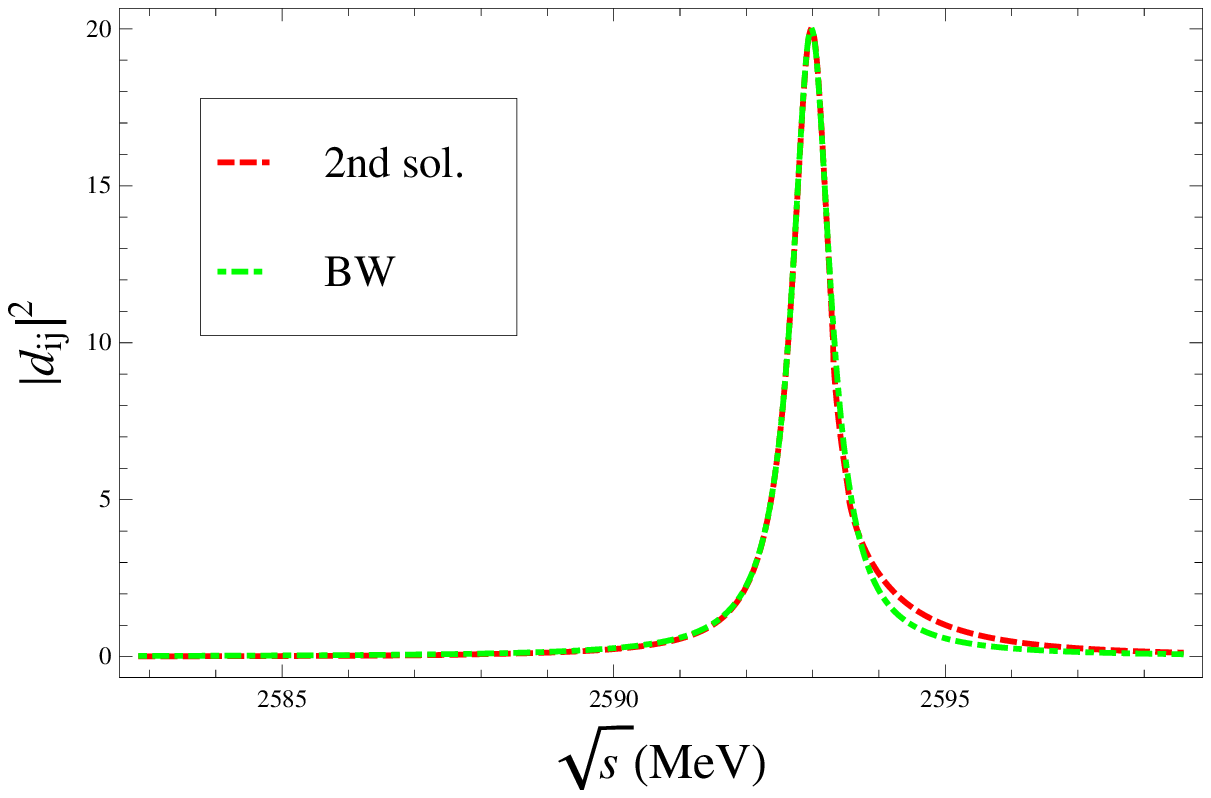}
\end{tabular}
\caption{{\protect{\small From left to right, the function $|d_{ij}(s)|^2$ is plotted along the physical axis for the 1st (blue solid line) and 2nd (red dashed line)  solutions
 for the coupled-channel case with stable $\Sigma_c$ when the $\Lambda_c^+(2595)$ pole is reproduced in the RS (1,1,1).
 We compare them with the standard Breit-Wigner formula (green dash-dotted line) with mass and width corresponding to the resulting poles in the
RS (1,0,0) for each solution: $2589.5-i\,0.3$~MeV  (1st sol.) and $2593.0-i\,0.4$~MeV (2nd sol.).
\label{fig.221015.1}}}}
\end{center}
\end{figure}

The presence of the CDD pole produces a rather strong distortion of the resonance signal in $t_{ij}(s)$ because of the nearby zero at $\sqrt{s}=M_{CDD}$, which is close to
the peak.
This is also the case for the resonance $ f_0(500)$  due to the strong distortion that the Adler zero, as required by chiral symmetry, produces in the scalar and isoscalar $\pi\pi$ scattering,
see e.g. Ref.~\cite{ref.221501.1.jao.04}.
Indeed, as also discussed in the previous reference, the production processes are not mediated by $t_{ij}(s)$ itself but by the so-called $d_{ij}(s)$ function, given in our case by
\begin{align}
\label{221015.1}
d_{ij}(s)=&t_{ij}(s)/{\cal K}_{ij}(s)=\big(1+{\cal K}_{ij}(s)\left[G_1(s)+G_2(s)+G_3(s)\right]\big)^{-1}~\,,
\end{align}
where the zero of $t_{ij}(s)$ caused by the CDD pole is removed. 
We plot  $|d_{ij}(s)|^2$ in Fig.~\ref{fig.221015.1} for the 1st (blue solid line) and 2nd solution (red dashed line) along the same energy interval as in the first two panels of Fig.~\ref{fig.211015.1}.
Due to the absence of the zero in $d_{ij}(s)$ associated with the CDD pole, one observes a resonance structure with a shape very close to that of a standard Breit-Wigner (BW) formula, $bw(\sqrt{s})$,
defined by
\begin{align}
bw(\sqrt{s})=&|d_{ij}(m_R)|^2\frac{\gamma_R^2/4}{(\sqrt{s}-m_R)^2+\gamma_R^2/4}~.
\label{171215.1}
\end{align}
Here $m_R$ represents the position of the maximum height of the resonance peak, denoted by $|d_{ij}(m_R)|^2$, and  $\gamma_R$ stands for its width.
The mass and width parameters in the BW formula are consistent with the pole positions in the RS (1,0,0) and the resulting curves are
 given by the green dash-dotted lines in Fig.~\ref{fig.221015.1}.

 By slightly changing the value of $M_{CDD}$ in the 1st-solution case, one can narrow down the gap between the pole in the (1,0,0) RS and the experimental one for the $\lamc$ in Eq.~\eqref{150915.1}.
 Our determination is $\mcdd = 2594.7(6)$~MeV and the pole position in the (1,0,0) RS then becomes $2592.4(4)- i 0.9(1)$~MeV.
 The mass obtained is compatible with the experimental value within uncertainty, while the width ($1.8\pm 0.2$~MeV) is still  slightly low, though compatible at the one-sigma level with Eq.~\eqref{150915.1}.
For the 2nd solution with $\mcdd = 2594.3(5)$~MeV  we find that  the mass of the pole in the (1,0,0) RS  is compatible at the one-sigma level with the experimental value
 but in all cases the width remains always much smaller than that of the $\Lambda_c^+(2595)$.  The values for $\mcdd$ reported here by considering the RS (1,0,0)
are gathered in the fifth column of Table~\ref{tab.091015.2}. 
In this respect, we consider the 1st solution as the preferred one in our study. These results are illustrated in Figs.~\ref{fig.tij100} and \ref{fig.dij100} where we plot the resulting
$|t_{ij}(s)|^2$ and $|d_{ij}(s)|^2$ for the 1st (blue solid lines) and 2nd (red dashed lines) solutions with these new values for $M_{CDD}$, respectively. From these figures it is clear that
the resonance structure for the 1st solution is much wider than that for the 2nd one, as well as compared with those plotted previously in Figs.~\ref{fig.211015.1} and \ref{fig.221015.1}. 
 For the 1st solution in the left panel of Fig.~\ref{fig.dij100} there is a departure from the BW shape in the tail to the right of the peak because of the opening of the next threshold, $\pi^+\Sigma_c^{0}$,
that precisely coincides with the starting energy of the shoulder.
 In the next section we include the widths of the $\Sigma_c$'s,  which are precisely as large as that of the $\Lambda_c^+(2595)$.
 It turns out that after taking this new physical effect into play, the agreement between the experimental width and the resulting one
for the 1st solution improves, while keeping a proper value for the mass, but still the width that stems from the 2nd solution is much smaller than the value in Eq.~\eqref{150915.1}.

\begin{figure}[H]
\begin{center}
\begin{tabular}{c}
  \includegraphics[width=.6\textwidth]{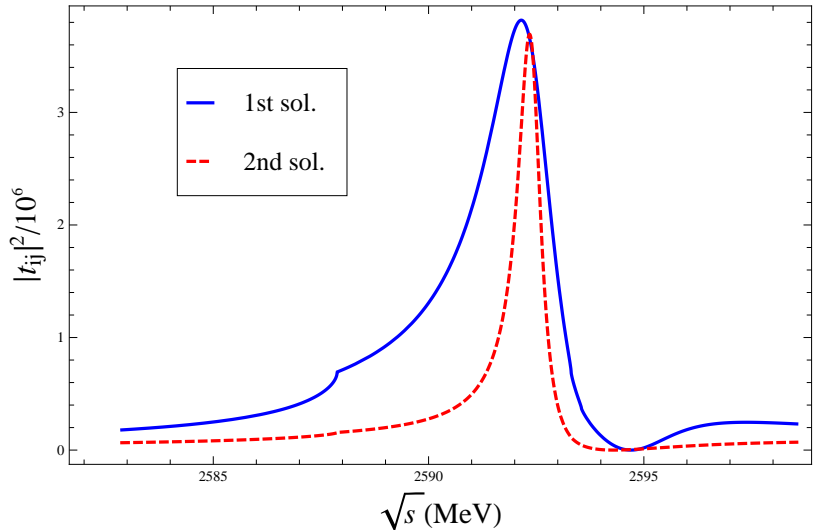}
\end{tabular}
\caption{{\protect{\small The partial wave $|t_{ij}(s)|^2$ for the coupled-channel case with stable asymptotic states
is plotted along the physical axis for the 1st (blue solid line) and 2nd (red dashed line)  solutions, respectively.  
The values of $\mcdd$ in the fifth column of Table~\ref{tab.091015.2} are used to plot  these curves.
See the text for details. 
\label{fig.tij100}}}}
\end{center}
\end{figure}

\begin{figure}[H]
\begin{center}
\begin{tabular}{cc}
  \includegraphics[width=.45\textwidth]{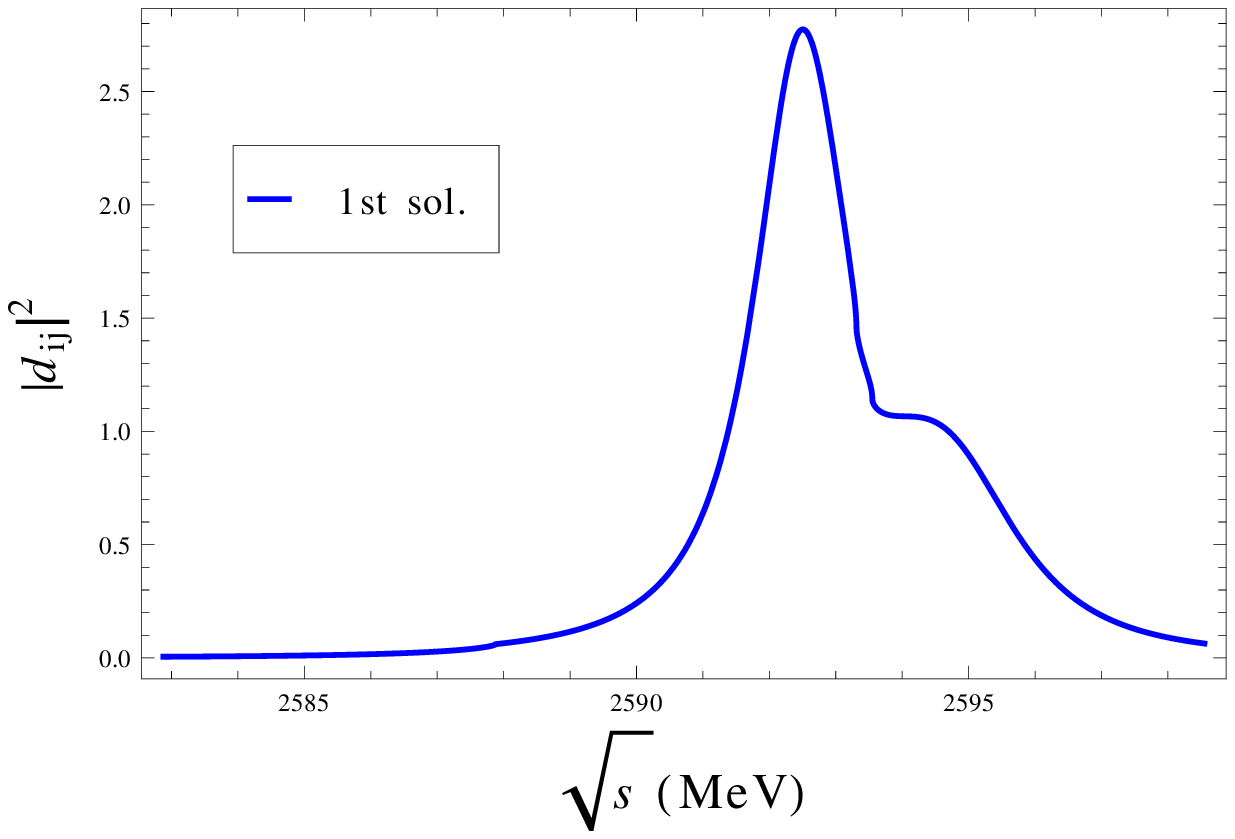}
 & \includegraphics[width=.45\textwidth]{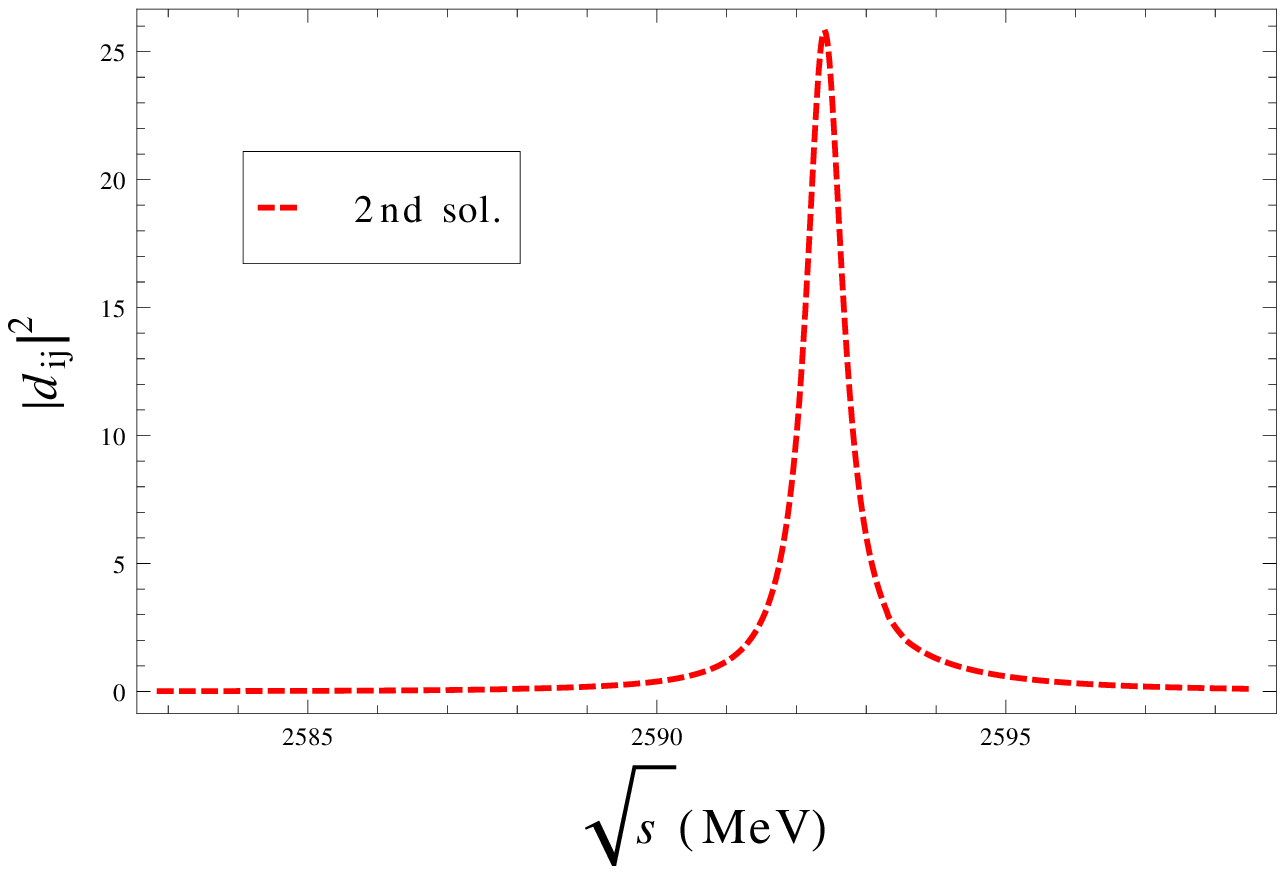}
\end{tabular}
\caption{{\protect{\small The function $|d_{ij}(s)|^2$ for the coupled-channel case with asymptotic stable states
is plotted along the physical axis for the 1st (blue solid line) and 2nd (red dashed line)  solutions, respectively.
 The values of $\mcdd$ in the fifth column of Table~\ref{tab.091015.2} are used. See the text for details.  \label{fig.dij100}}}}
\end{center}
\end{figure}


\section{Coupled-channel CDD analysis II: including the widths of $\Sigma_c$}
\label{sec.211015.1}

Now we estimate the effects by taking into account the small but finite widths of the $\Sigma_c$ baryons to evaluate $t(s)$
from Eq.~\eqref{201015.1}. As indicated above we take for the widths of $\Sigma_c^{++}$ and $\Sigma_c^0$ the central values
 provided by the updated PDG \cite{150915.1.pdg}, $\Gamma_{\Sigma_c^{++}}=1.89$~MeV and $\Gamma_{\Sigma_c^0}=1.83$~MeV.   For
the $\Sigma_c^+$ we use in the following the value $\Gamma_{\Sigma_c^+}=1.8$~MeV, because it is naturally expected  that its decay width 
 should be saturated also by the strong decay to $\Lambda_c^+\pi$~\cite{150915.1.pdg} and then its value should be very close to the widths
 of its other isospin multiplet companions. Note also that the decay channel $\Lambda_c^+\pi$ has a much lighter threshold than the $\Sigma_c$ mass,
so that there is  plenty of phase space available and then the width should be quite insensitive to
small changes in the mass of the decaying particle.

\begin{figure}[H]
\begin{center}
\begin{tabular}{cc}
  \includegraphics[width=.475\textwidth]{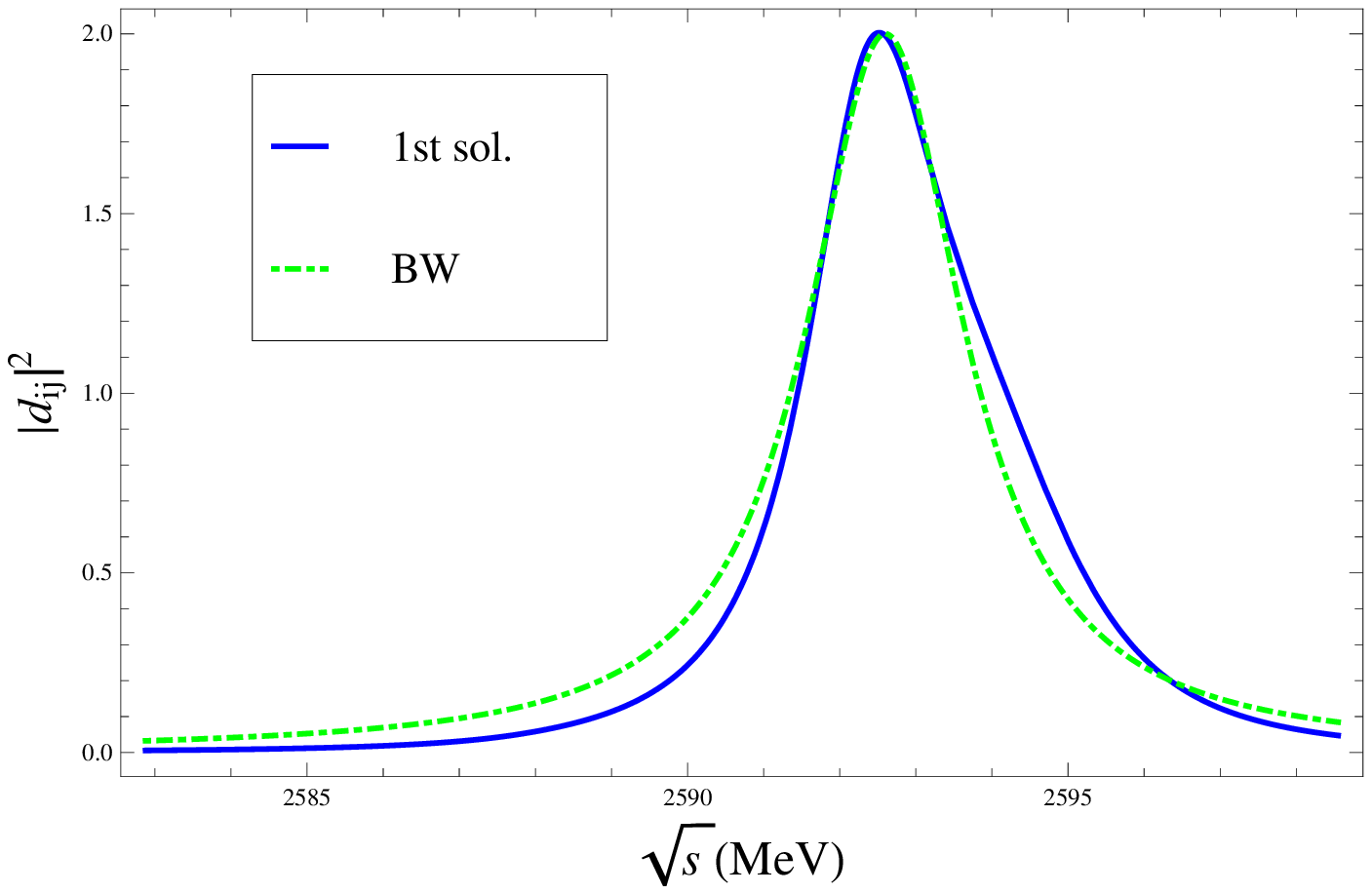} &
  \includegraphics[width=.475\textwidth]{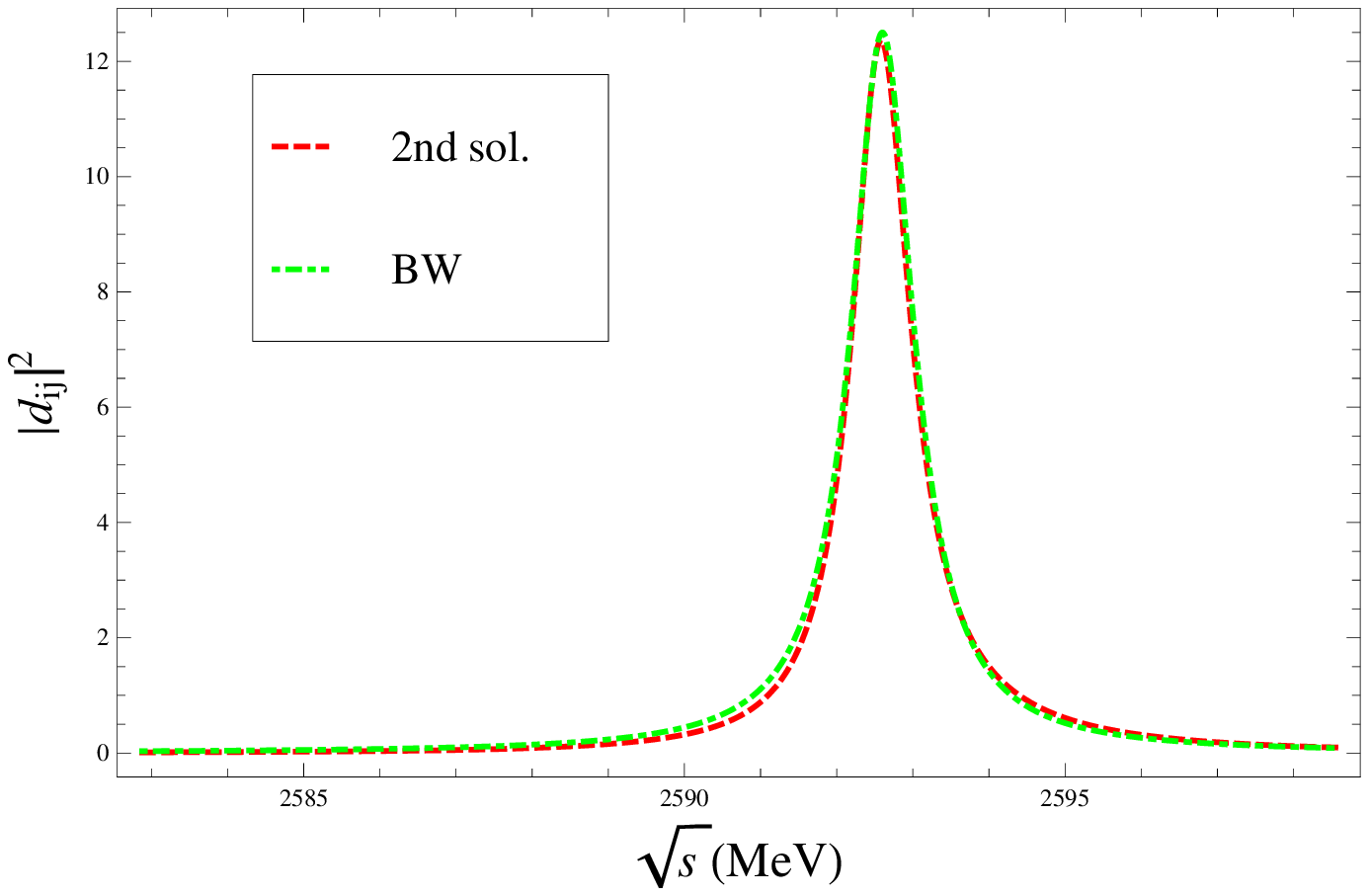}
\end{tabular}
\caption{{\protect{\small The function $|d_{ij}(s)|^2$  is plotted along the physical axis for the coupled-channel case including the finite widths of  $\Sigma_c$.
For the meaning of the lines see Fig.~\ref{fig.221015.1} and the text in Sec.~\ref{sec.211015.1}. 
\label{fig.221015.2}}}}
\end{center}
\end{figure}

  We follow the formalism of Ref.~\cite{131015.1.jao.10} to take into account the $\Sigma_c$ widths in Eq.~\eqref{201015.1},
 which consists of employing complex masses for the $\Sigma_c$, with the replacement $M_{\Sigma_c} \to M_{\Sigma_c}-i\Gamma_{\Sigma_c}/2$.
In this way,  when evaluating $t_{ij}(s)$, given by Eq.~\eqref{201015.1}, the changes only affect the unitarity loop functions
 $G_i(s)$, while ${\cal K}_{ij}(s)$ in Eq.~\eqref{141015.3}, is not changed.

As a technical remark we briefly discuss now how to perform the analytical extrapolation into the unphysical RS's when the finite widths
of the $\Sigma_c$ are considered. We take the nonrelativistic limit to simplify expressions and for channel $j)$ after including  the finite widths Eq.~\eqref{220915.1} transforms into
\begin{align}
\label{031115.1}
k_j=&\sqrt{2\bar{\mu}_j}\sqrt{\sqrt{s}-M_{j\Sigma_c}-M_{j\pi}+i\Gamma_{j\Sigma_c}/2}~,\nn\\
\bar{\mu}_j=&\frac{M_{j\pi}(M_{j\Sigma_c}-i\Gamma_{j\Sigma_c}/2)}{M_{j\pi}+M_{j\Sigma_c}-i\Gamma_{j\Sigma_c}/2}~,
\end{align}
with $M_{j\pi}$ and $M_{j\Sigma_c}$ the masses of the pion and $\Sigma_c$ corresponding to channel $j$, in order.
  According to Eq.~\eqref{031115.1}  $k_j$ has a complex branch point at $M_{j\pi}+M_{j\Sigma_c}-i\Gamma_{j\Sigma_c}/2$, with a horizontal cut running to the right from this
singularity in the  complex  $\sqrt{s}$ plane. The transition to the associated 2nd RS when crossing this cut downwards is obtained by the replacement $k_j \to -k_j$ and then
 one can move deeper in the lower half plane of this RS. In this way the RS (1,0,0) is obtained by following
 this procedure only for the lightest threshold ($\pi^0\Sigma_c^+$), while the RS (1,1,1) would require
to apply it to all the three channels.

We are now in a position to look for  the pole in the RS (1,0,0) that connects continuously to the physical axis in the energy region around the $\lamc$ mass and is responsible for the resonance signal of the $\Lambda_c^+(2595)$.
 As in Sec.~\ref{sec.091015.1},  we take the values of $\alpha$ and $\gamma$ for every solution in Table~\ref{tab.091015.2} and readjust $M_{CDD}$
so that the resulting pole in the RS (1,0,0) is inside the energy region corresponding to the mass of the $ \Lambda_c^+(2595)$ resonance, Eq.~\eqref{150915.1}.
The resulting value of $\mcdd$ for each solution is given in the last column of Table~\ref{tab.091015.2}, around 2594~MeV. This value is indeed almost coincident with
that  obtained already in the case of stable $\Sigma_c$ (5th column of Table~\ref{tab.091015.2}), and perfectly compatible within errors.
   For the 1st solution we find a pole at the position  $(2592.3\pm 0.3)-i \, (1.13 \pm 0.15)$~MeV, that reproduces very well
 the parameters for the $\Lambda_c^+(2595)$ resonance.
 However, for the 2nd solution although we can easily get the correct mass  the width is always much smaller than the experimental value,
 with the pole located at $(2592.6\pm 0.4)-i \, (0.47 \pm 0.05)$~MeV. The mass ($M_R$) and width ($\Gamma_R$) corresponding to each of these poles are also given in
Table~\ref{tab.091015.2}. 
 We see that the changes both in $\mcdd$ and pole positions are small compared with the case of stable asymptotic states,
 which is a  welcome stability in the results and conclusions. Nevertheless, the increase in the width between 
 a $10\%$ to $20\%$ 
   for the 1st solution makes its central value well inside the one-sigma level of the experimental value in Eq.~\eqref{150915.1}.

 Next, we plot  $|d_{ij}(s)|^2$ in Fig.~\ref{fig.221015.2} by using the central values of the parameters, 
where the left panel is for the 1st solution (blue solid line) and the right one for the 2nd solution (red dashed line).
 We further estimate the mass and width of the resonance signal by comparing $|d_{ij}(s)|^2$ with the BW formula, Eq.~\eqref{171215.1},
which is drawn in Fig.~\ref{fig.221015.2} with the green dash-dotted lines.
 For the 1st solution the resulting BW resonance parameters are $m_R=2592.6(5)$ MeV  and $\gamma_R=2.5(5)$~MeV and for the
 2nd case one has $m_R=2592.6(5)$ MeV  and $\gamma_R=1.0(2)$~MeV.
 It is clear that  the resulting pole positions for the 1st and 2nd solutions (columns 6 and 7 in Table~\ref{tab.091015.2}) agree quite
closely with the BW parameters.  This implies again that the pole responsible  for the resonance
signal of $\lamc$ lies in RS (1,0,0), rather than in the RS (1,1,1).
 The BW parameters reflect once more  that while the 1st solution is able to give the correct resonance signal
corresponding to the experimental parameters for the $\Lambda_c^+(2595)$, both mass and width,
the 2nd solution is not able to reproduce the width, which is  less than 50\% of the experimental one.
This is another good reason to disfavor the 2nd solution within our analysis. The shoulder above the $\pi^+\Sigma_c^0$ threshold
clearly present in Fig.~\ref{fig.dij100} has now almost disappeared in Fig.~\ref{fig.221015.2} due to the finite widths of the $\Sigma_c$. 
 The latter dilutes the threshold effects, which are displaced off the real axis into the complex plane as discussed above.

\begin{table}[ht]
\begin{center}
\begin{tabular}{lcccc}
\hline
                      & $\alpha$          & $\gamma$\,[MeV]         & $M^{i)}_{CDD}$\,[MeV] &  $M^{iii)}_{CDD}$\,[MeV] \\
\hline
1st solution       & $-0.03422(4)$   & $1.9(1)$      & $2592.6(1)$        & $2587.8\pm 1.1$           \\
2nd solution       & $-0.03387(6)$   & $2.5(3)$      & $2593.4(3)$        & $2598.3\pm 0.8$          \\
\hline
\end{tabular}
{\caption {\small From left to right, values for the parameters $\alpha$, $\gamma$, $M_{CDD}^{i)}$ and $M_{CDD}^{iii)}$ 
after imposing that $t(s)$ of Eq.~\eqref{021015.1} has a pole at $s_R$ for channels $i)$ and $iii)$ by including the finite-width effects of the $\Sigma_c$.
\label{tab.sol.width}}}
\end{center}
\end{table}

It is important to remark that our conclusions are also stable if we redetermine the  
parameters $\alpha$ and $\gamma$ in Table~\ref{tab.091015.1} by repeating the same single-channel analysis as in Sec.~\ref{sec.021015.1} but with finite widths of $\Sigma_c$ included.
The new values obtained for the parameters  are given in Table~\ref{tab.sol.width}.
It is clear that  $\alpha$ and $\gamma$ from the 1st solution are quite stable when taking the finite-width effects into account,
while for the 2nd one the changes
 in these parameters are larger, specially for $\gamma$ that decreases from
$3.0$ to $2.5$~MeV, though the numbers are still compatible within errors.
 We have also explicitly verified that if one uses the values of $\alpha$
 and $\gamma$ from Table~\ref{tab.sol.width}  in the coupled-channel
analysis with nonzero widths for the $\Sigma_c$  the conclusions do not change.
Namely, the 1st solution still provides a pole for the $\Lambda_c^+(2595)$ that is
compatible with experiment while the pole for the 2nd solution has a width around
a factor 2 smaller than the experimental one.
 Nevertheless,  we prefer to present the
detailed analysis for the  results of $\alpha$ and $\gamma$ from Table~\ref{tab.091015.1}, instead of those from Table~\ref{tab.sol.width}, because 
in this way we can more clearly identify the finite-width effects, since
the same input values for $\alpha$ and $\gamma$ are used both in the zero- and finite-width cases.

Another way to account for the finite-width effects is to perform a convolution of the $G(s)$ function with a spectral mass 
distribution by considering the $\Sigma_c$ width~\cite{Roca:2006sz,131015.1.jao.10}
\begin{eqnarray}\label{eq.covg} 
\widetilde{G}(s,M_\pi,M_{\Sigma_c})=\frac{1}{N}\int_{M_{\Sigma_c}-2\Gamma_{\Sigma_c}}^{M_{\Sigma_c}+2\Gamma_{\Sigma_c}}
d\sqrt{s'}\,\, {\rm Im}\bigg[ \frac{1}{\sqrt{s'}-M_{\Sigma_c} + i\Gamma_{\Sigma_c}/2}\bigg] G(s,M_\pi,\sqrt{s'}) \,,
\end{eqnarray}
with the function $G(s)$ given in Eq.~\eqref{300915.2} and the normalization factor $N$ corresponding to 
\begin{eqnarray}
N= \int_{M_{\Sigma_c}-2\Gamma_{\Sigma_c}}^{M_{\Sigma_c}+2\Gamma_{\Sigma_c}}
d\sqrt{s'}\,\, {\rm Im}\bigg[ \frac{1}{\sqrt{s'}-M_{\Sigma_c} + i\Gamma_{\Sigma_c}/2}\bigg]  \,.
\end{eqnarray}
In order to clearly show the differences among the results of zero-width and those with finite-width for the $\Sigma_c$, evaluated 
either with complex masses or  making the convolution of $G(s)$, 
we study the $d(s)$ function of the single-channel case from the three scenarios. In this case the function $d(s)$ reads
\begin{align}
\label{231015.1}
d(s)=&\left[1+\omega(s) G(s)\right]^{-1}\,,
\end{align}
where three different scenarios are distinguished through the $G(s)$ functions. 
 We plot in Fig.~\ref{fig.231015.1} the modulus squared of $|d(s)|^2$
on the physical real axis with the $\pi\Sigma_c$ masses corresponding to $\pi^0\Sigma_c^+$, case $i)$. The same values for the parameters $\alpha$, $\gamma$ and $\mcdd^{i)}$ in Table~\ref{tab.091015.1} are used when plotting the curves in Fig.~\ref{fig.231015.1}, where the left (right) panel is for the 1st (2nd) solution. 
In this way the  differences between  lines in every panel are purely caused by the way how the $\Sigma_c$ widths are implemented. 
 We consider the cases with zero width for $\Sigma_c^+$ (blue solid lines), as originally done in Sec.~\ref{sec.021015.1},
 with a complex mass for this baryon (red dashed lines), $M_{\Sigma_c^+}\to M_{\Sigma_c^+}-i \Gamma_{\Sigma_c^+}/2$, 
and with the spectral convolution of the $G(s)$ function in Eq.~\eqref{eq.covg} (green dot-dashed lines). 
We see that the changes are small for both solutions and the results are stable, with also little changes in the resulting pole positions.
While for the coupled-channel case, if we take the same input values for $\alpha$, $\gamma$ and $\mcdd$ in the scenarios with complex masses
 and convolution of  $G(s)$  with a spectral mass distribution, visible differences appear in the curves.
Nevertheless, by slightly changing the value of $\mcdd$ within 0.5~MeV, we can easily get similar results for these two cases in the 1st solution.
For the 2nd solution, though the heights of $|d(s)|^2$ are somewhat different, 
it is easy to obtain similar positions of the peaks. Therefore, 
the results and conclusions obtained by using the function $G(s)$ convoluted with a spectral mass distribution are not changed compared to the case of
 using complex masses. Thus,  the 1st solution is able to reproduce the resonance parameters of the $\lamc$, while the 2nd one
 is not. 

\begin{figure}[H]
\begin{center}
\begin{tabular}{cc}
  \includegraphics[width=.475\textwidth]{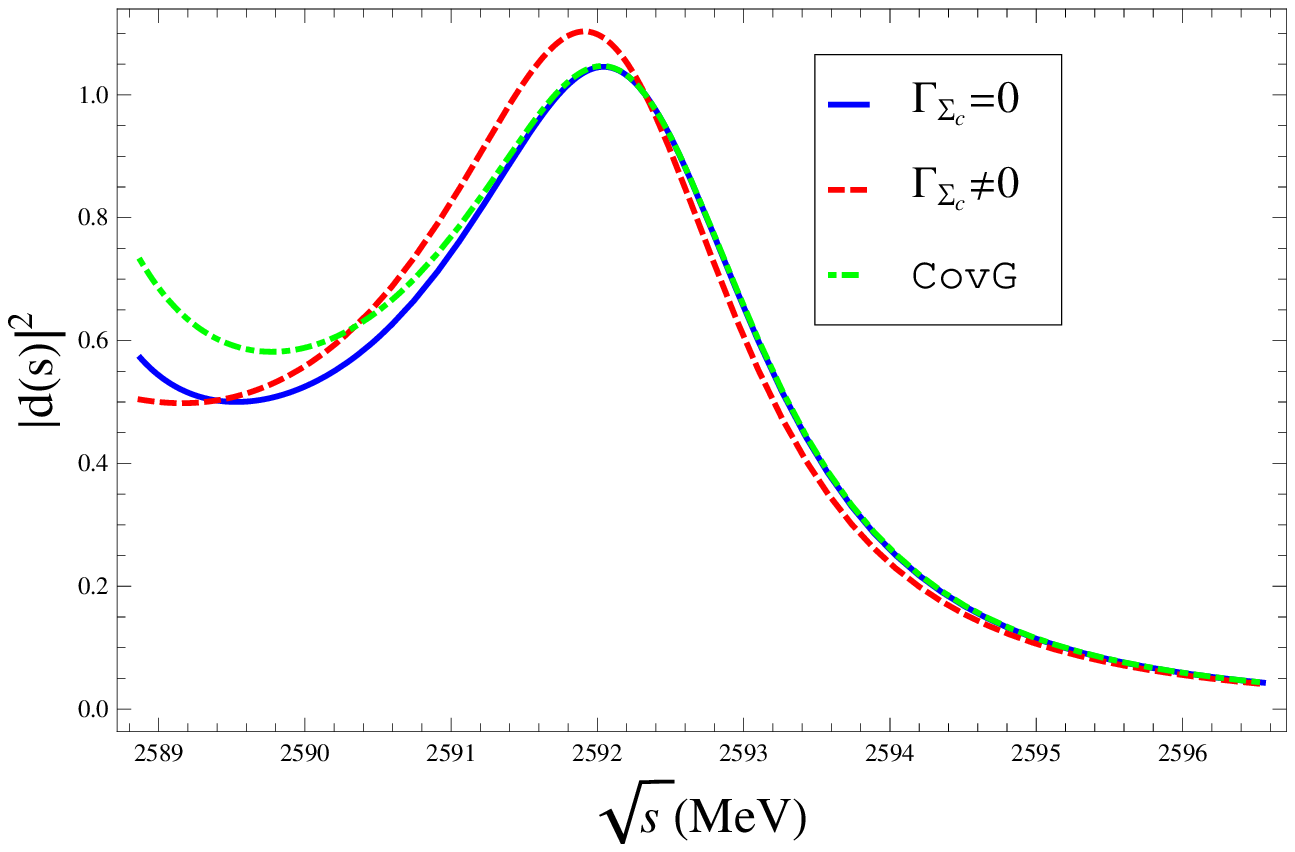} &
  \includegraphics[width=.475\textwidth]{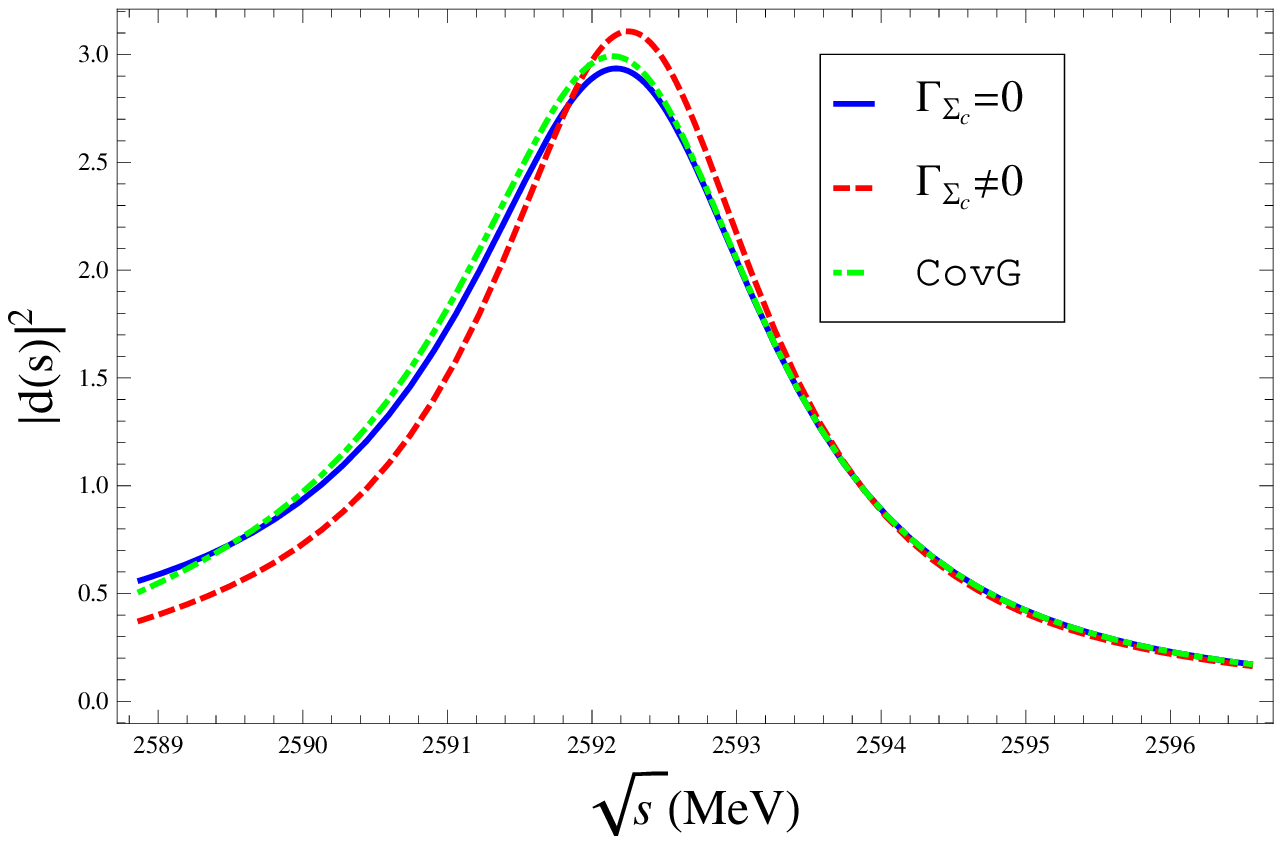}
\end{tabular}
\caption{{\protect{\small The function $|d(s)|^2$ in the single-channel case  is plotted along the physical axis.
The same values for the parameters $\alpha, \gamma, \mcdd^{i)}$ from Table~\ref{tab.091015.1} are used for all the curves.  The blue solid, red dashed, green dot-dashed lines stand 
for the results with zero $\Sigma_c^+$ width, finite $\Sigma_c^+$ width using a complex mass and the function $G(s)$ convoluted, respectively.
The left panel is for the 1st solution and the right panel is for the 2nd one. }
\label{fig.231015.1}}}
\end{center}
\end{figure}

\section{Compositeness study of the $\Lambda_c^+(2595)$ }
\label{sec.151015.1}

The presence of a nearby CDD pole to the $\Lambda_c^+(2595)$  resonance mass, with $\left| M_{CDD}-M_R\right| \approx \Gamma_R$, as it follows
from Tables~\ref{tab.091015.1} and \ref{tab.091015.2},  is a clear indication  that there is an important component, maybe dominant one,
due to non $\pi\Sigma_c$ degrees of freedom, either corresponding to other channels not included, e.g. heavier ones like $DN, D^*N, ...$~\cite{Hofmann:2005sw,Geng:2014ina,Haidenbauer:2010ch,Ramos:2006vq,Ramos:2009vq,Nieves:2008dp,Nieves:2012hm,Nieves:2015jsa,Liang:2014kra,Lutz:2003jw}, or to the quark and gluon compact states~\cite{Copley:1979wj,Zhong:2007gp,Pirjol:1997nh,Tawfiq:1998nk,Zhu:2000py,Blechman:2003mq,Migura:2006ep}.

In order to quantify this statement we apply here  the theory developed in Ref.~\cite{151015.1.guo.15} that allows
a probabilistic interpretation of the compositeness relation \cite{Weinberg:1962hj,Baru:2003qq,ref.230715.4,ref.230715.5,ref.230715.6,Agadjanov:2014ana}
for resonances under the condition that $\sqrt{{\rm Re}s_R}$ is larger than the lightest threshold. According to the Ref.~\cite{151015.1.guo.15}  the weight of
an open  two-body channel $j$ to the resonance compositeness, $X_j$, is given by
\begin{align}
\label{161015.1}
X_j=|g_j|^2\left|\frac{\partial G_j(s_R)}{\partial s}\right|~,
\end{align}
with $g_j^2$ the residue of $t(s)$ to channel $j$ at the resonance pole position $s_R$,
\begin{align}
\label{161015.2}
\lim_{s\to s_R}(s-s_R)t_{jj}(s)=-g_j^2~.
\end{align}
The difference between 1 and the sum of  $X_j$ over the open channels considered is the elementariness $Z$,
 which measures the weight of all other components in the resonance.

We first apply Eq.~\eqref{161015.1} to the single-channel study of Sec.~\ref{sec.021015.1} for case $i)$, since then the criterion of applicability of Eq.~\eqref{161015.1} is fulfilled as $M_{\pi^0}+M_{\Sigma_c^+}<\sqrt{{\rm Re}s_R}$, with the resonance lying in the RS that connects continuously with the physical axis above this threshold. We then obtain the values of $X$ given in the 2nd column of Table~\ref{tab.231015.1}
for the 1st and 2nd solutions, in order from top to bottom, with the input parameters taken from Table~\ref{tab.091015.1}.
We also give the absolute value of the residue of $t(s)$ at $s_R$, $|g|^2$ in the third column of Table~\ref{tab.231015.1}. 
It is clear that $X$ turns out to be small indicating that, as expected, the non-$\pi\Sigma_c$ components are  dominant,
so that $Z>0.8$ holds for both solutions.

 The results of Ref.~\cite{151015.1.guo.15} can also be applied to the channel $i)$ in the case of the coupled-channel analysis
for stable asymptotic states.  The results that follow for $X_1$ and $|g^2|$ are given in the last two columns  of Table~\ref{tab.231015.1},
respectively.\footnote{In this table we have multiplied by 3 the residue squared to compensate for the 1/3 introduced in  Eq.~\eqref{141015.3}, so that
the comparison with the single-channel case is more straightforward.}
 Notice that here $X_1$ is the compositeness coefficient for channel 1 only, while in the single-channel analysis $X$ corresponds to all three $\pi\Sigma_c$ channels.
 For the preferred 1st solution the value of $X_1$ is smaller than $X$ but significantly larger than $X/3$,
 while for the 2nd solution $X_1$ is much smaller than $X$ but 3$X_1$ is similar to the latter.
 These results clearly indicate that the $ \pi^0\Sigma_c^+$ channel has a small contribution to the composition of the $\lamc$.
In turn, the value of $X$ for the single-channel case suggests that the $(\pi \Sigma_c)^+$ total component in this resonance is also small, although
 we cannot be conclusive here since the isospin breaking effects could distort the values of $X_2$ and $X_3$ (which are the compositeness coefficients
 of $\pi^+\Sigma_c^0$ and $\pi^-\Sigma_c^{++}$, respectively) 
from $X/3$. We simply cannot exclude that this could be the case.

\begin{table}
\begin{center}
\begin{tabular}{rcccc}
\hline
  &   $X$    & $|g|^2\,$[GeV$^2$]  &  $X_1$ & $ 3 |g|^2\,$[GeV$^2$]  \\
\hline
  1st sol.        &   $0.14 \pm 0.02$   &  $13.0 \pm 1.8$     &   $0.11 \pm 0.02$      &   $29.9 \pm 3.9$     \\
  2nd sol.        &   $0.17 \pm 0.04$    &  $15.2 \pm 2.7$     &   $0.04 \pm 0.01$     &  $9.8 \pm 1.4$    \\
\hline
\end{tabular}
\end{center}
\caption{{\protect{\small Values for the compositeness $X$ and residue $|g|^2$: single-channel
case (2nd and 3rd columns); poles in the (1,0,0) RS for the $3\times 3$ coupled-channel analysis with zero $\Sigma_c$ widths (4th and 5th
columns).
\label{tab.231015.1}}}}
\end{table}

These non-$\pi\Sigma_c$ components in $\Lambda_c(2595)$, to which the CDD pole is associated,
 could correspond to heavier channels, like $DN, D^*N,$ etc, as proposed in 
 Refs~\cite{Hofmann:2005sw,Geng:2014ina,Haidenbauer:2010ch,Ramos:2006vq,Ramos:2009vq,Nieves:2008dp,Nieves:2012hm,Nieves:2015jsa,Liang:2014kra,Lutz:2003jw},
three-body $\pi\pi\Lambda_c$ (in connection with the coupled-channel study of Sec.~\ref{sec.211015.1} including the finite width for the $\Sigma_c$ so as
 to reproduce more accurately the $\lamc$ width), as well as to possible more elementary degrees of freedom from the QCD Lagrangian (quarks and gluons),
 as discussed in Refs.~\cite{Copley:1979wj,Zhong:2007gp,Pirjol:1997nh,Tawfiq:1998nk,Zhu:2000py,Blechman:2003mq,Migura:2006ep}.
Having obtained that $X_1\simeq 0.10$ and $X\simeq 0.15$, cf. 1st solution of Table~\ref{tab.231015.1}, suggests also that the contributions from the two-body $(\pi\Sigma_c)$ states
 are small but still noticeable in the resonance composition \cite{Geng:2014ina}.

\section{Conclusions}
\label{sec:sum}

In this paper we develop a general framework that goes beyond effective range expansion
 to scrutinize the situation with a resonance pole locating very close to the underlying thresholds.
In particular we apply this formalism to make a thorough and delicate study of the $\lamc$, which just lies between the $\pi^0\Sigma_c^+$ and $\pi^+\Sigma_c^0, \pi^-\Sigma_c^{++}$ thresholds.

We show that in order to give the correct $\lamc$ pole in the effective range expansion of the single-channel or uncoupled scattering
 one needs  large magnitudes for the scattering length ($a$)
and effective range ($r$). The latter could have an absolute value as large as 40~fm, which certainly indicates the presence of an extra and small
 energy scale beyond the natural range for strong forces ($\sim 1$~fm).
 Moreover the values of $a$ and $r$ are extremely sensitive to the actual masses used for the isospin multiplet $\pi\Sigma_c$.
E.g. the value of $a$ resulting in the $\pi^0\Sigma_c^+$ channel compared to the one for the other two channels changes around one order of magnitude.
 This is indicative of a highly striking fine-tuned physical scenario.
 We then develop a formalism that is applicable in the nearby threshold region with typical three-momenta involved much smaller than $m_\pi$. 
It is  based on the general form that partial-wave amplitudes have when only right-hand cuts are present, and it is then applied to both the single- and coupled-channel cases.  
In particular, the striking phenomena just referred are linked and could be
 naturally explained by the presence of a CDD pole near the $(\pi\Sigma_c)^+$ thresholds, which also prevents
the application of effective range expansion up to the pole position of the $\lamc$ resonance.
 In the coupled-channel formalism, we find that the resonance signal showing up
in the real physical axis corresponds to the pole appearing in the 2nd Riemann sheet, instead of the 3rd one. This finding is further quantified
 by comparing Breit-Wigner functions with our outputs for the line shapes on the real axis, so that the Breit-Wigner mass and width
correspond closely to the pole position in the 2nd Riemann sheet.

The finite-width effects from the $\Sigma_c$ are studied too. It is shown that these contributions, although  typically small,  improve
 the description of the $\lamc$ pole,  so that the 1st solution can properly reproduce the experimental values for the $\lamc$ pole position.
 However, for the 2nd solution,
though the mass is well reproduced, the resulting width is always a factor 2 smaller than the experimental value.
  Therefore the 1st solution is considered to be the favored one in our study (it is also the one that provides the most stable results
when passing from the zero- to the finite-width $\Sigma_c$ analyses).
 Finally, we make the compositeness analysis of $\pi \Sigma_c$  for the $\lamc$ and our result in this respect is that the compositeness of $\pi^0\Sigma_c^+$
 inside $\lamc$ is neatly small ($\lesssim 10\%$). This result, together with the crucial role of the CDD pole near the $\pi\Sigma_c$ threshold,
 indicates that non-$\pi\Sigma_c$ degrees of freedom are essential in the $\lamc$ resonance, e.g. heavier hadronic channels, such as
$DN, D^*N$, or compact quark-gluon structures, are likely to be the dominant components inside the $\lamc$.

We foresee that the formalism developed here could be useful in other similar systems,
 such as the exotic $XYZ$ heavy-flavor states~\cite{150915.1.pdg}.

\subsection*{Acknowledgments}
 This work is supported in part
by the MINECO (Spain) and ERDF (European Commission) grant FPA2013-40483-P and the Spanish
 Excellence Network on Hadronic Physics with contract No.~FIS2014-57026-REDT, 
 the National Natural Science Foundation of China (NSFC) under Grant Nos.~11575052 and 11105038, the Natural Science Foundation of Hebei Province with contract No.~A2015205205,
the grants from the Education Department of Hebei Province under contract No.~YQ2014034,
the grants from the Department of Human Resources and Social Security of Hebei Province with contract No.~C201400323.


\end{document}